\documentclass[a4paper,12pt]{article}

\title{\LARGE\bf SLE, CFT and zig-zag probabilities.}
\date{}
\author{}

\newcommand{\monh}{\mathbb{H}}
\newcommand{\monht}{\mathbb{H}_t}
\newcommand{\monK}{\mathbb{K}}
\newcommand{\monu}{\mathbb{U}}
\newcommand{\monv}{\mathbb{V}}

\newcommand{\tild}{\widetilde}

\newcommand{\ket}[1]{| #1 \rangle}
\newcommand{\bra}[1]{\langle #1 |}
\newcommand{\aver}[1]{\langle \omega | #1 | \omega \rangle}
\newcommand{\statav}[1]{\langle #1 \rangle}

\def\debut{\begin{eqnarray}}
\def\fin{\end{eqnarray}}

\usepackage[final]{graphicx}
\usepackage{amssymb} 
\usepackage{oldgerm} 

\begin{document}
\maketitle

\vspace{-1.5 truecm}

\centerline{\large Michel Bauer\footnote{Email:
    bauer@spht.saclay.cea.fr} and Denis Bernard\footnote{Member of
    the CNRS; email: dbernard@spht.saclay.cea.fr}} 

\vspace{.3cm}

\centerline{\large Service de Physique Th\'eorique de Saclay}
\centerline{CEA/DSM/SPhT, Unit\'e de recherche associ\'ee au CNRS
\footnote{URA 2306 du CNRS}}
\centerline{CEA-Saclay, 91191 Gif-sur-Yvette, France}



\vspace{1.0 cm}

\begin{abstract}
  The aim of these notes is threefold. First, we discuss geometrical
  aspects of conformal covariance in stochastic Schramm-Loewner
  evolutions (SLEs). This leads us to introduce new ``dipolar'' SLEs,
  besides the known chordal, radial or annular SLEs. Second, we review
  the main features of our approach connecting SLEs to conformal field
  theories (CFTs). It is based on using boundary CFTs to probe the SLE
  hulls.  Finally, we study zig-zag probabilities and their relation
  with CFT correlation functions. We suggest a putative link between
  the braiding of SLE samples and that of CFT correlation functions.
\end{abstract}


\vskip 1.5 truecm


\section{Introduction.}

The purpose of these notes \footnote{These notes are written for the
  proceedings of the NATO conference on \textit{Conformal Invariance
    and Random Spatial Processes} held in Edinburgh in July 2003.} is
to give a compact exposition of the way we have been thinking about
the relationships between stochastic Schramm-Loewner evolution (SLE)
and conformal field theory (CFT). Part of the material is an abridged,
but hopefully readable, version of previously published work. Our
approach consists in using CFT to probe SLE hulls.  However, we have
tried to put more emphasis on general features. The other part
contains some new remarks.  In the short description that follows, we
stress the original points that do not appear elsewhere in our
previous work.

In the first part, we look at the relationship between stochastic
Loewner evolution (SLE) and conformal field theory (CFT) with emphasis
on conformal covariance.  We stress the role of two vector fields that
underly the geometry of conformally covariant stochastic differential
equations. The possibility to embed the Lie algebra they generate in a
manageable Lie sub-algebra of the Virasoro algebra is at the heart of
the SLE-CFT relationship.  We treat from this unified geometric
viewpoint the definition of SLE and its connection with CFT under two
symmetry conditions : the domain and the boundary conditions for SLE
should be invariant under a one parameter group of conformal
automorphisms and under an involution.  Apart from the well studied
chordal and radial cases, these conditions are met in two cases,
dipolar and annular SLE. We have not met the dipolar case in the
previous literature. It has the property that dipolar SLE hulls never
touch a forbidden interval on the boundary.  The annular case involves a
simple moduli space. These two cases have not yet been investigated
thoroughly.

Next we recall how a special representation of the Virasoro algebra,
degenerate at level 2, allows to construct a martingale generating
function for SLE and we explain the CFT physical picture that
underlies this construction. We give details for the chordal case, and
brief discussions of the other, slightly more involved, three cases. A
short discussion of the explicit implementation of conformal
transformations in CFT follows, together with its applications to the
Virasoro Wick theorem and to representation theory. This is then used
to define CFTs in the complement of the SLE hulls which is the key
point for coupling CFTs to SLEs.

The purpose of the second part is to illustrate, in the example of
chordal SLE, how to perform probabilistic SLE computations in the CFT
framework. We review briefly the CFT interpretation of the restriction
martingale as a CFT partition function. We then turn to boundary and
bulk properties of the hulls by relating the probability that the SLE
trace zig-zags in a prescribed order through a finite number of small
balls (small intervals on the boundary, small disks in the bulk) to an
explicit correlation function in CFT. After a review of previously
computed functions, we treat some new manageable cases explicitly. We
also present a CFT computation of the fractal dimension of the SLE
trace. The conformal operators involved in all these probabilities, as
well as their dimensions, are explicited. However to completely
specify the value of the correlation functions, the intermediate
states (or conformal blocks) that propagate from one operator to the
next have to be determined.  For that we do not have a general
answer and we have to rely on a case by case analysis using the
appropriate boundary conditions. Nevertheless, the answer to this well
posed problem contains important data because, due to the extended
structure of the hulls, the bulk correlation functions depend on the
order of visits which means that some non trivial CFT monodromies,
alias braiding relations, quantum group actions and yangians should
appear naturally in SLEs.

\section{A few avatars of SLE.}

 From a general point of view, stochastic Schramm-Loewner evolutions
\cite{schramm0} describe, by means of a stochastic differential
equation, the growth of random curves or hulls emerging from the
boundary of a Riemann surface. One of the crucial features should be
conformal covariance, in the sense that if two Riemann surfaces are
related by a conformal map $\varphi$, respective SLE samples should be
related by $\varphi$ as well.

Although interesting attempts have been made \cite{makarov,
  zhang,friedkalk}, the implementation of SLE in such a general
setting is still in its infancy. The situations that have been
thoroughly studied up to now is when the Riemann surface is a simply
connected domain $\mathbb{U}$ in the complex plane and the hulls
either connect two points on the boundary $\partial \mathbb{U}$
(chordal SLE) or one point on the boundary and one in the bulk of
$\mathbb{U}$ (radial SLE), see
refs.\cite{schramm0,RhodeSchramm,LSW,Lawler}. In the following, we
shall try to abstract common features of these two cases and use these
to look at two little known examples.

\subsection{Conformal covariance.}
\label{sec:confcov}

Our starting point is a discussion of conformal covariance for
stochastic differential equations in the following sense. It is well
known that trajectories of points on manifolds are related to vector
fields. The case of interest for us is when the manifold is a Riemann
surface $\Sigma$. Suppose $z \in \monu \subset \mathbb{C}$ is a
coordinate system for some open subset of $\Sigma$ and $\varphi$ maps
$\monu$ conformally to some domain $\monv \subset \mathbb{C}$. Suppose
that an intrinsic motion of points on $\Sigma$, when written in the
local coordinate in $\monu$, satisfies the family of differential
equations $df_t(z)=dt\, \nu(f_t(z))$, with initial conditions
$f_0(z)=z$, where $\nu$ is holomorphic in $\monu$. Then, when written
in the local coordinate in $\monv$, the corresponding map is
$f^\varphi_t\equiv \varphi \circ f_t \circ \varphi^{-1}$, which
satisfies $df^\varphi_t=dt\, \nu^\varphi \circ f^\varphi_t$ with
$\nu^\varphi \circ \varphi = \varphi' \nu$.  This equation expresses
that $w=\nu(z)\partial_z$ is a holomorphic vector field on some open
subset of $\Sigma$.

What happens now if the motion on $\Sigma$ is stochastic ? Suppose
that $\xi_t$ a Brownian motion with covariance ${\bf
  E}[\xi_t\,\xi_s]=\kappa\, {\rm min}(t,s)$ and that the motion, 
read in $\monu$, satisfies
\begin{eqnarray}
 \label{eqdiff}
df_t(z)=dt\, \sigma(f_t(z))+ d\xi_t \, \rho(f_t(z)),  
\end{eqnarray} 
 in the following sense :
for each trajectory $\xi_t$ there is a random but almost surely positive
time $T$ and a nonempty open domain $\monu_T \subset \monu$ such that
$f_t$ maps $\monu_T$ into $\monu$ and solves the above differential
equation for $t \in [0,T]$ and $z \in \monu_T$. Then, by It\^o's
formula, the motion read in $\monv$ satisfies $df^\varphi_t=dt\,
\sigma^\varphi \circ f^\varphi_t + d\xi_t \,\rho^\varphi\circ
f^\varphi_t$ with $f^\varphi_t\equiv \varphi \circ f_t \circ
\varphi^{-1}$, and
\begin{equation}
  \label{eq:transf}
\rho^\varphi \circ \varphi = \varphi' \rho \quad \mathrm{and} \quad
\sigma^\varphi \circ 
\varphi =\varphi' \sigma+\frac{\kappa}{2} \varphi'' \rho^2.
\end{equation}
By a simple rearrangement,  this means that 
\begin{equation}
\label{eq:rel}
w_{-1} \equiv \rho(z)\, \partial_z \quad \mathrm{and} 
\quad w_{-2}\equiv \frac{1}{2}\left(-\sigma(z)+
  \frac{\kappa}{2}\rho(z)\rho'(z)\right)\partial_z
\end{equation}
are holomorphic vector fields on an open subset of $\Sigma$.  Under
the motion, some points may leave this open subset before time $t$.
The corresponding random subsets of $\monu$ and $\monv$ are
related by $\varphi$.

To make contact with group theory, let us assume that \textit{i)}
there is a linear space $O$ of holomorphic functions and a group $N$
that (anti) acts faithfully on $O$ by composition $\gamma_f \cdot F
\equiv F \circ f$ for $F \in O$ and $f \in N$, and \textit{ii)} that
$f_t \in N$, at least up to a possibly random but strictly positive
time.  In this situation we may view $f_t$ as a random process on
$N$. For chordal and radial SLE we shall exhibit appropriate $O$ and
$N$ later.

It\^o's formula shows that $\gamma_{f _t}^{-1} \cdot d \gamma_{f _t} .
\cdot F =(dt \, \sigma + d\xi_t \, \rho)F'+ dt
\,\frac{\kappa}{2}\rho^2 F''$, or equivalently
\begin{eqnarray}
\gamma_{f _t}^{-1} \cdot d \gamma_{f _t}=dt \,
(-2w_{-2}+\frac{\kappa}{2}w_{-1}^2)- d\xi_t \, w_{-1}.
\label{SLEcov}
\end{eqnarray}
This equation involves only intrinsic geometric objects. It is at the
heart of the relation between SLE and conformal field theory.
Eq.(\ref{SLEcov}) can be transformed into an ordinary differential
equation for  $g_t \equiv e^{\xi_t w_{-1}}\cdot f_t$,  i.e. 
$\gamma_{g_t}=\gamma_{f _t}e^{\xi_t w_{-1}}$,  which  satisfies
$\gamma_{g_t}^{-1} \cdot d \gamma_{g_t}=-2dt \, (e^{-\xi_t
w_{-1}}w_{-2}e^{\xi_t w_{-1}})$.

The structure of the Lie algebra generated by $w_{-1}$ and $w_{-2}$
plays an important role, and the possibility to embed this Lie algebra
in the Virasoro algebra is crucial. Recall that the Virasoro algebra
$\mathfrak{vir}$ is the Lie algebra generated by $\{L_n, \; n\in
\mathbb{Z},\; c\}$ with commutation relations :
\begin{equation}
[L_n,L_m] = (n-m)L_{n+m} +\frac{c}{12}(n^3-n) \delta_{n+m,0} \qquad
[c,L_n]=0.
\label{viral}
\end{equation}
It is the (essentially unique) central extension of the Lie algebra
of Laurent polynomial vector fields in $\mathbb{C}$ via the map
$\ell_n=-z^{n+1}\partial_z \rightarrow L_n$.  The relevant properties of
$\mathfrak{vir}$ and its representations will be presented in section
\ref{sec:CFTchorSLE}.

\subsection{Chordal SLEs.}
\label{sec:chordal}

Chordal SLE \cite{schramm0,RhodeSchramm,LSW,Lawler}
describes the local random growth of hulls between two
points on the boundary of a simply connected domain in $\mathbb{C}$.
Let us start with the definition when
the domain is the upper half plane and the hulls grow from the
origin to the point at infinity.

A hull in the upper half plane $\monh =\{z \in \mathbb{C}, \Im {\rm
  m}\, z > 0\}$ is a bounded subset $\monK \subset \monh$ such that
$\monh \setminus \monK$ is open, connected and simply connected.
The local growth of a family of hulls $\monK_t$ parameterized by $t \in
[0,T[$ with $\monK_0=\emptyset$ is related to complex analysis as
follows. By the Riemann mapping theorem, $\monht\equiv \monh \setminus
\monK_t$, the complement of $\monK_t$ in $\monh$, which is simply
connected by hypothesis, is conformally equivalent to $\monh$ via a
map $g_t$.  This map can be normalized to behave as $g_t(z)
=z+2t/z+O(1/z^2)$, using the $PSL_2(\mathbb{R})$ automorphism group
of $\monh$ and an appropriate time parameterization.  The crucial
condition of \textit{local} growth leads to 
the Loewner differential equation
\begin{eqnarray}
\partial_t g_t(z)=\frac{2}{g_t(z)-\xi_t}\ ,\quad g_{t=0}(z)=z
\label{loew}
\end{eqnarray}
with $\xi_t$ a real function. For fixed $z$, $g_t(z)$ is
well-defined up to the time $\tau_z$ for which
$g_{\tau_z}(z)=\xi_{\tau_z}$.  

Chordal stochastic Loewner evolution is obtained \cite{schramm0} by
choosing $\xi_t=\sqrt{\kappa}\, B_t$ with $B_t$ a normalized Brownian
motion and $\kappa$ a real positive parameter so that
${\bf E}[\xi_t\,\xi_s]=\kappa\,{\rm min}(t,s)$.  The SLE hull is
reconstructed from $g_t$ by $\monK_t=\{z\in{\monh}:\ \tau_z\leq
t\}$ and the SLE trace $\gamma_{[0,t]}$ by
$\gamma(t)=\lim_{\epsilon\to0^+}g_t^{-1}(\xi_t+i\epsilon)$.  Basic
properties of the SLE hulls and SLE traces are described in
\cite{schramm0,RhodeSchramm,LSW}.  In particular, $\gamma_{[0,t]}$ is
almost surely a curve.  It is non-self intersecting and it coincides
with $\monK_t$ for $0<\kappa\leq 4$, while for $4<\kappa<8$ it
possesses double-points and it does not coincide with $\monK_t$. For
$8\leq \kappa$ it is space filling. The normalization at infinity
ensures that (almost surely) $\gamma_t \rightarrow \infty$ when
$t\rightarrow \infty$.

To make contact with stochastic differential equations and conformal
covariance, it is useful to define $f_t(z)
\equiv g_t(z) - \xi_t$ which satisfies 
$$ 
d f_t = \frac{2dt}{f_t}-d\xi_t.
$$

According to our previous discussion, $w_{-1}=-\partial_z=\ell_{-1}$
and $w_2=-\frac{1}{z}\partial_z=\ell_{-2}$. The first vector field is
holomorphic in $\monh$ and tangent to the boundary, so that by the
Schwarz reflection principle it extends automatically to a holomorphic
vector field in the full complex plane. The second one is holomorphic
in $\monh$ and tangent to the boundary except at the origin. For
the same reason it extends automatically to a holomorphic vector field
in the complex plane with the origin removed ; the extension has a
simple pole with residue $2$ as its sole singularity. Both $w_{-1}$
and $w_{-2}$ vanish at infinity, a double and triple zero
respectively, which accounts for the fact that $f_t(z)=z +O(1)$ at
infinity. They have no other common zero, which is the geometric
reason why the SLE trace goes to infinity at large $t$. The germ of
$f_t$ at infinity belongs to the group $N_-$ of germs of holomorphic
functions at $\infty$ of the form $z+\sum_{m \leq -1} f_{m} z^{m+1}$
(with real coefficients).  The group $N_-$ (anti)acts by composition
on $O_-$, the space of germs of holomorphic functions at $\infty$
fixing $\infty$. In particular, to $f_t$ we can associate
$\gamma_{f_t}\in N_-$, which satisfies eq.(\ref{SLEcov}) i.e.
explicitly
\begin{eqnarray}
\gamma_{f_t}^{-1} \cdot d \gamma_{f_t}=dt(\frac{2}{z}\partial_z
+\frac{\kappa}{2}\partial_z^2)-d\xi_t\partial_z.
\label{itoN-moi}
\end{eqnarray} 

We shall see later that physically meaningful representations of the
Virasoro algebra can be turned into representations of $N_-$. This is
closely related to the fact that the Lie algebra generated by
$-\partial_z$ and $-\frac{1}{z}\partial_z$ is isomorphic to the
negative (locally) nilpotent sub-algebra of the Virasoro algebra.

If one thinks in terms of singularities, the most general holomorphic
vector field in $\monh$ vanishing at infinity and tangent to the
boundary is a linear combination of $\ell_{-1}$ and $\ell_0=-z\partial_z$.
Similarly, the most general holomorphic vector field in $\monh$
vanishing at infinity and tangent to the boundary but at the origin,
where it has a simple pole of residue $2$, is $-2\ell_2$ modulo a linear
combination of $\ell_{-1}$ and $\ell_0$. Imposing left-right
symmetry leads to $w_{-1}\propto \ell_{-1}$ and $w_{-2}-\ell_{-2} \propto
\ell_0$. The $\ell_0$ contribution can be eliminated by a rescaling of $f_t$
and a deterministic time change preserving the Brownian character of
$\xi_t$. Note that to define Brownian motion along a curve, one
needs a parameterization. The fact that $\ell_{-1}$ is the infinitesimal
generator of a one parameter group of conformal automorphisms of
$\monh$ that extend to the boundary can be viewed as providing such a
parameterization. 

Here are again some remarks on conformal covariance.

If $\monu$ is a general simply connected domain
(possibly with some regularity conditions for the boundary) with two
marked points $P_0$ and $P_{\infty}$ on the boundary, the Riemann
mapping theorem can be applied to map the complement of a hull in
$\monu$ to $\monu$ itself just as before. Let $\varphi$ be the
conformal transformation, well defined up to an irrelevant dilation in
$\monh$, that maps $\monh$ to $\monu$, $0$ to $P_0$ and $\infty$ to
$P_{\infty}$. The uniformizing map for the hull $\varphi(\monK)$ in
$\monu$ is $\varphi \circ f \circ \varphi^{-1}$ where $f$ is the
uniformizing map for the hull $\monK$ in $\monh$. The Loewner equation
for growing hulls in $\monu$, say in the normalization when the tip of
the hull is always mapped to $P_0$, then follows by transport. 

One can reformulate this as follows. Consider the abstract Riemann
surface $\Sigma$ with two marked boundary points which is the equivalence class
of all open subsets $\monu$ of $\mathbb{C}$ conformally equivalent to
$\monh$ with $P_0$ and $P_\infty$ mapped to $0$ and $\infty$. The
above discussion means that the Loewner equation and the Loewner hulls
are intrinsic geometric objects on $\Sigma$.  From the point of view
of probabilities, this is not really a big constraint : once a
probability measure on hulls in $\monh$ is defined, one can always
transport it to the other non empty simply connected open sets in
$\mathbb{C}$. It is plain that the transported measure is also related
to Loewner uniformization when the one in $\monh$ is.

What is highly nontrivial is that continuum limits of discrete 2D
critical statistical mechanics models are conformally covariant. Such
models are usually defined on a lattice, say
$a\mathbb{Z}+ia\mathbb{Z}$ where $a > 0$ is a unit of length.
Criticality is the statement that when $a$ goes to zero certain
nontrivial physical observables survive and do not depend on any
scale. The limit has to be defined carefully. If $\monu$ and $\monv$
are two conformally equivalent open subsets of $\mathbb{C}$, related
by a conformal map $\varphi$, one can consider the model in the
intersection of the lattice $a\mathbb{Z}+ia\mathbb{Z}$ with $\monu$ or
$\monv$. When $a$ goes to $0$, scale invariance does not a priori
imply that the limit theories on $\monu$ and $\monv$ are related in a
simple way. Based on heuristic arguments of locality, it was
conjectured in \cite{BPZ} about twenty years ago, using another
language, that the limiting theory is well defined on the
abstract Riemann surface which is the equivalence class of all open
subsets of $\mathbb{C}$ conformally equivalent to $\monu$. For
instance, as recalled in section \ref{sec:conftrans}, correlation
functions of local observables become sections of appropriate bundles,
i.e.  have transformations that involve derivatives of $\varphi$ when
going from $\monu$ to $\monv$. Interfaces are directly related by
$\varphi$ and the probability law governing their fluctuations as well.
So SLE hulls behave geometrically as they should to encode the
statistics of critical interfaces. However, SLE does it in a very
specific way, involving Loewner evolution and $1d$ Brownian motion. Even
if, as we shall recall in section \ref{sec:CFTchorSLE}, there is a
simple and elegant direct relation between SLE and CFT, we have no
rigorous understanding why the fluctuations of critical interfaces
are so closely related to the Markov property and the continuity which
characterize Brownian motion. At least intuitively, conformal invariance is
related to the Markov property of SLE conformal maps and the fact
that interfaces are not supposed to branch is related to continuity. 
 
We end this section with a caveat. The above statement, that SLE is
conformally covariant under domain change, should not be confused with
the, \textit{incorrect in general}, statement that SLE growth
processes are conformally invariant in the sense $2d$ Brownian motion
is conformally invariant. A local conformal transformation maps
Brownian motion to Brownian motion modulo a random time change. But
the classical locality computation, see e.g. ref.\cite{Lawler}, shows
that SLE is conformally invariant in that sense only for $\kappa=6$,
which corresponds to percolation, for which the central charge and the
conformal weight to be introduced later both vanish.

\subsection{Radial SLEs.}

Radial SLE \cite{schramm0,RhodeSchramm,LSW,Lawler}
describes the growth of a hull from a point on
the boundary of a simply connected domain $\mathbb{U}$ in the complex
plane to a point inside $\mathbb{U}$. Conformal covariance and
conformal automorphisms allow to choose $\monh$ as
$\mathbb{U}$, $0$ as the boundary point where the SLE trace emerges
and $i$ as the inside point where the SLE trace converges. In terms of
geometry of vector fields, we can still use holomorphicity, the Schwarz
symmetry principle and left-right symmetry, and impose that $-2w_{-2}$
has after extension a simple pole at the origin with residue $2$
(this is just a choice of time scale, the crucial point
is that it has positive residue), and that $w_{-1}$ is
holomorphic. The sole difference is that this time the vector fields
have to vanish at $i$, where the SLE trace converges. This gives two
real conditions, so the situation is more rigid than in the chordal
case. One finds  $w_{-2}=-\frac{(1+z^2)}{z}\partial_z$, and
$w_{-1}\propto -(1+z^2)\partial_z$. The choice of the proportionality
factor is just a normalization of $\kappa$.
For the space $O$ we choose this time the germs of holomorphic
functions at $i$ fixing $i$ and $N$ is the subspace of $O$ made of the
germs with non vanishing derivative at $i$. Eq.(\ref{SLEcov}) reads
explicitly 
\begin{eqnarray}
\gamma_{f_t}^{-1} \cdot d \gamma_{f_t}=dt\left(2\frac{1+z^2}{z}\partial_z
+\frac{\kappa}{2}((1+z^2)\partial_z)^2\right)-d\xi_t(1+z^2)\partial_z.
\label{itoNi}
\end{eqnarray}
Observe that this time we do not use translations but another one
parameter subgroup of the group of conformal automorphisms of $\monh$,
namely the ones fixing $i$, to parameterize the real axis and define
Brownian motion.

Contact with conformal field theory is via the equalities
$$
w_{-2}=\ell_{-2}+\ell_0 \quad w_{-1}=\ell_{-1}+\ell_1.$$
Notice that
$$e^{-\xi_t
w_{-1}}w_{-2}e^{\xi_t w_{-1}}=-(1+z^2)\frac{1+z\tan \xi_t}{z-\tan
\xi_t} $$
so that the ordinary differential equation governing radial SLE in
$\monh$ is
$$\partial_t g_t(z)=2(1+g_t(z)^2)\frac{1+g_t(z)\tan \xi_t}{g_t(z)-\tan
\xi_t}. $$

The more traditional presentation of radial SLE has the unit disk
$\mathbb{D}$ as $\mathbb{U}$, $1$ as the boundary point where the SLE
trace emerges and $0$ as the inside point where the SLE trace
converges. The disk variable is $Z=-\frac{z-i}{z+i}$. The vector field
$w_{-1}=-2iZ\partial_Z$ generates rigid rotations. Rescaling things to
eliminate the unesthetic factor of $2$, one retrieves the standard
radial stochastic Loewner equation. With this normalization, one finds
$w_{-1}=-iZ\partial_Z$ and $2w_{-2}=Z\frac{Z+1}{Z-1}$. Notice for
further use that for a holomorphic vector field $v(Z)\partial_Z$ in
the unit disk tangent at the boundary, the Schwarz symmetry principle
reads $v(1/\overline{Z})=-\overline{v(Z)}/\;\overline{Z}^2$, which is
of course satisfied by $w_{-1}$ and $w_{-2}$.

\subsection{Dipolar SLEs.}

If one realizes that radial SLE is closely linked to a compact Cartan
torus of $SL_2(\mathbb{R})$, related to rigid rotations of the disk,
it is tempting to look at what we call dipolar SLE, obtained by
replacing rotations by a non compact Cartan torus of
$SL_2(\mathbb{R})$. This amounts to replace the complex fixed point
$i$ by the pair of real fixed points $1$ and $-1$ and leads to
$$ w_{-2}=\ell_{-2}-\ell_0=-\frac{(1-z^2)}{z}\partial_z \quad
w_{-1}=\ell_{-1}-\ell_1=(1-z^2)\partial_z,$$
For $O$ and $N$, one has two natural choices : germs of holomorphic
functions at $\pm 1$ fixing $\pm 1$.
One can check that the corresponding ordinary differential equation,
$$\partial_t g_t(z)=2(1-g_t(z)^2)\frac{1-g_t(z)\tanh
  \xi_t}{g_t(z)-\tanh \xi_t}. $$
is the Loewner equation when the
Loewner map is normalized to fix $1$ and $-1$ and have the same derivative
at these two points : $g'_t(\pm 1)=e^{-4t}$.

The simple time independent case, obtained by replacing the Brownian
$\xi_t$ by a constant $\xi$, can be solved explicitly. For $\xi=0$
the solution is $g_t^{(0)}(z)=\sqrt{e^{-4t}z^2+1-e^{-4t}}$ and the
SLE trace is $\gamma_t=i\sqrt{e^{4t}-1}$ which swallows the positive
imaginary axis. For general $\xi$, $g_t=h_{\xi}\circ
g_t^{(0)}\circ h_{\xi}^{-1}$ where $h_{\xi}(z)=\frac{z+\tanh
  \xi}{1+z\tanh \xi}$ is the conformal transformation fixing $\pm1$
and mapping $0$ to $\tanh \xi$. The trace is the semi circle starting
at $\tanh \xi$ and ending at $\coth \xi$.

\begin{figure}[htbp]
  \begin{center}
    \includegraphics[width=0.9\textwidth]{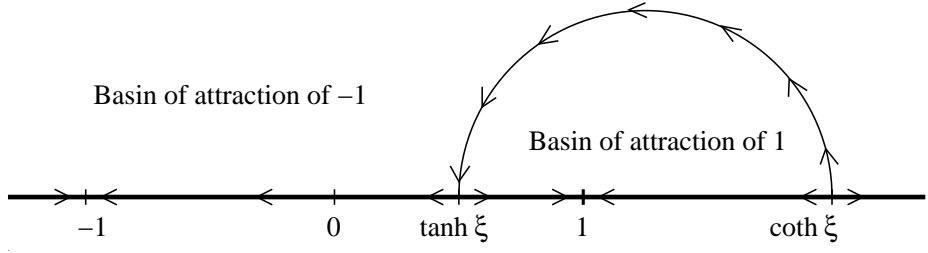}
      \caption{\em Qualitative picture of the dipolar 
Loewner vector field.}
     \label{fig:dipolar}
 \end{center}
\end{figure}

The vector field $2(1-z^2)\frac{1-z\tanh \xi}{z-\tanh \xi}\partial_z$
vanishes at three points, $\pm 1$ and $\coth \xi$ and has a simple
pole at $\tanh \xi$. For positive $\xi$, it looks as in Figure
\ref{fig:dipolar}.  The fixed points $\pm 1$ are attractive, but
$\coth \xi$ is repulsive. The structure of the vector field at the
pole illustrates the square root singularity of the Loewner map at the
tip of the trace.  The trace is the unstable curve, joining the pole
to the repulsive fixed point, that separates the basins of attraction
of $1$ and $-1$. It takes an infinite time to reach its end point.

Let us return to the case when $\xi_t$ is a Brownian motion.  Since
the complement $I$ of $]-1,1[$ in $\overline{\mathbb{R}}$ is invariant
under the SLE map, a simple property is that the hull cannot hit $I$.
A rough argument goes as follows. Assume the contrary. By continuity,
before the hull hits $I$, it has to leave the half disk $|z|<1/2$. The
probability $P_{1/2}(t)$ that this happens before time $t$ is $<1$ for
every $t$ and increases with $t$. The probability $P_I(t)$ that the
hull hits $I$ before time $t$ has the same properties. Now, the time
when the hull leaves the half disk $|z|<1/2$ is a stopping time, say
$\tau_{1/2}$.  For $t \geq \tau_{1/2}$, $f_t\circ f_{\tau_{1/2}}^{-1}$
is a new SLE process, starting from $0$. As $I$ is invariant under the
time evolution, the original process hits $I$ only if the new one
does.  Independence yields $P_I(t)=\int_0^t P_I(t-t') dP_{1/2}(t')
\leq P_I(t) P_{1/2}(t)$ from which we deduce that $P_I(t)=0$ because
$P_{1/2}(t)<1$.

Together with the static picture, this leads to the following
qualitative picture for dipolar SLE traces.  The Brownian motion will
oscillate indefinitely, little by little covering the whole real line.
As $\coth \xi_t$ approaches $\pm 1$ exponentially fast for large
$\xi_t$, a plausible scenario is that most of the time the SLE trace
will move towards points close to $\pm 1$, hesitating alternatively
between the two points, but never swallowing $\pm1$.  Dipolar SLE
certainly deserves a rigorous investigation.

\subsection{Annular SLEs.}

There is yet another geometry with all the useful properties used in
the above examples, namely the possibility to apply the Schwarz
symmetry principle, left-right symmetry and a one parameter group of
automorphisms. The geometry is that of an annulus. One new feature is
the existence of moduli. This case was treated from another viewpoint
by \cite{zhang}. The case of percolation $\kappa=6$ has been
investigated by \cite{dubedat}.

It is a classical theorem that any open domain in $\mathbb{C}$ with
the topology of an annulus is conformally equivalent to an open
annulus $\mathbb{A}_p=\{Z \in \mathbb{C}, e^{-p}<|Z|<1\}$ for some
unique $p \in ]0,\infty[$. The automorphism group of $\mathbb{A}_p$ is
generated by rigid rotations and the discrete symmetry $Z \rightarrow
e^{-p}/Z$ which exchanges the boundary components.

\begin{figure}[htbp]
  \begin{center}
    \includegraphics[width=0.9\textwidth]{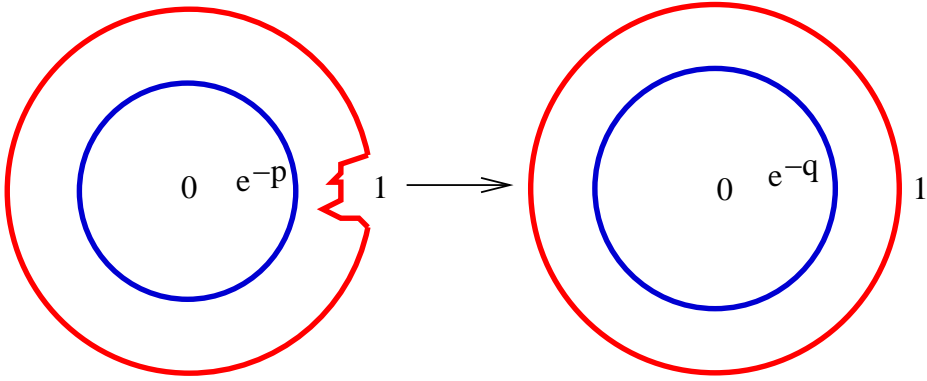}
      \caption{\em Uniformizing map.}
     \label{fig:annu1}
 \end{center}
\end{figure}

Suppose that a hull $\monK$ is removed from $\mathbb{A}_p$ and that
the boundary of the resulting domain contains the internal circle
$|Z|=e^{-p}$ and a non empty (open) interval of the external circle
$|Z|=1$.  The domain $\mathbb{A}_p \setminus \monK$ is conformally
equivalent to say $\mathbb{A}_q$ via a map $g$, well defined modulo
rotations, mapping the circle $|Z|=e^{-p}$ to the circle $|Z|=e^{-q}$,
see Figure(\ref{fig:annu1}).

\begin{figure}[htbp]
  \begin{center}
    \includegraphics[width=0.75\textwidth]{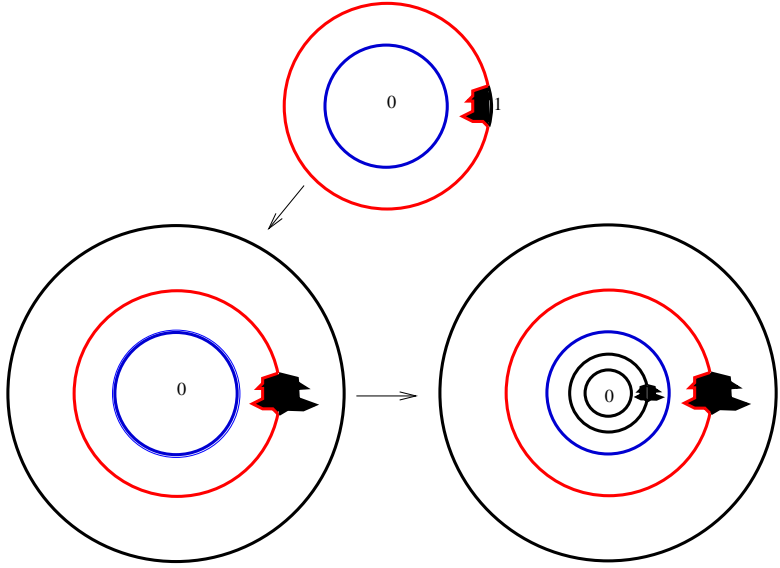}
      \caption{\em Recursive construction of $\mathbb{C}^*(p,\monK)$
        and extension of the holomorphy domain by
        successive Schwarz symmetries.}
     \label{fig:annu}
 \end{center}
\end{figure}

Let $G_p$ be the group acting on $\mathbb{C}^*$ generated by the two
transformations $S_0:Z\rightarrow 1/\overline{Z}$ and
$S_p:Z\rightarrow e^{-2p}/\overline{Z}$. As abstract groups, $G_p$ and
$G_q$ are isomorphic via $S_0 \leftrightarrow S_0$,
$S_p\leftrightarrow S_q$~: we write $T_p$ for a generic element in
$G_p$ and $T_q$ for the corresponding element of $G_q$. Note that
$\mathbb{A}_q$ is a fundamental domain for the action of $G_q$ on
$\mathbb{C}^*$. Let $\mathbb{C}^*(p,\monK)\equiv \mathbb{C}^*\setminus
\cup_{T_p\in G_p}T_p(\overline{\monK})$ be the connected open subset
of $\mathbb{C}^*$ obtained by removal of the $G_p$-translates of the
closure of $\monK$, see Figure(\ref{fig:annu}).

The Schwarz symmetry principle implies that $g$
can be extended to a holomorphic function on $\mathbb{C}^*(p,\monK)$
by $g(T_p(Z))=T_q(g(Z))$.
This extension satisfies
\begin{equation}
  \label{eq:g}
g(1/\overline{Z})=1/\overline{g(Z)}, \qquad
g(e^{-2p}/\overline{Z})=e^{-2q}/\overline{g(Z)}, \quad Z\in
\mathbb{C}^*(p,\monK).  
\end{equation}
The relationship between $p$, $q$ and $\monK$ is non
trivial.

In the infinitesimal case, when $\monK$ is a compact strict subset of
the circle $|Z|=1$, we write $g(Z)=Z+\epsilon v(Z)$ for some
holomorphic vector field on $\mathbb{C}^*(p,\monK)$ and $q=p-\epsilon
s$.  Eqs.(\ref{eq:g}) for $g$ translate into
\begin{equation}
  \label{eq:v}
v(1/\overline{Z})=-\overline{v(Z)}/\overline{Z}^2 \qquad e^{2p}
v(e^{-2p}/\overline{Z})=-\overline{v(Z)}/\overline{Z}^2+2s/\overline{Z}  
\end{equation}
for the vector field $v$.

To
describe the case of an infinitesimal slit emerging from $1$, we
impose that $\monK=\{1\}$ and $v(Z)\sim -\frac{2}{Z-1}$ for $Z \rightarrow
1$. A lengthy but elementary computation shows that the above system
has no solution if $s\neq 1$ and that if $s=1$ the most general
solution is an automorphic form~:
$$v(Z)/Z= 1+i\lambda-2\left(\sum_{m\geq 0} \frac{Z}{Z-e^{2mp}}+
  \sum_{m< 0}\frac{e^{2mp}}{Z-e^{2mp}}\right),$$ 
where $\lambda$ is an arbitrary real parameter. 
As usual, $w_{-1}$ is the generator associated to the
$\lambda$-ambiguity and imposing left-right symmetry leads to
$$w_{-1}=-iZ\partial_Z \qquad
w_{-2}^{[p]}=\left(\frac{Z(Z+1)}{2(Z-1)}+\sum_{m> 0}
  \frac{Z^2}{Z-e^{2mp}}+ \sum_{m<
    0}\frac{Ze^{2mp}}{Z-e^{2mp}}\right)\partial_Z.$$
One can check that $w_{-2}^{[p]}$ vanishes at $Z=-1$, as suggested by symmetry.

Several remarks are in order. The vector field $w_{-1}=-iZ\partial_Z$
is the same as for radial SLE. A new feature is the dependence of
$w_{-2}$ on the moduli $p$. When $p\rightarrow +\infty$, one recovers
radial SLE.  As expected, $s$ is not an independent parameter but is
determined by $\monK$. The value $s=1$ that we found implies that if
we iterate infinitesimal slits to make a Loewner evolution, the
parameter of the annulus varies linearly : $q=p-t$ with the
normalizations we have chosen. This leads to a slight modification of
eq.(\ref{eqdiff}) :
$$
df_t(z)=dt\, \sigma^{[p-t]}(f_t(z))+ d\xi_t \, \rho(f_t(z))  
$$
with $\sigma^{[p]}$ and $\rho$ related to $w_{-1}$ and $w_{-2}$ as in
eq.(\ref{eq:rel}). Standard arguments lead to the existence of
local (in time and space) solutions of this equation with initial data
$f_0(z)=z$. 
Correspondingly, eq.(\ref{SLEcov}) becomes~:
$$
\gamma_{f _t}^{-1} \cdot d \gamma_{f _t}=dt \,
(-2w_{-2}^{[p-t]}+\frac{\kappa}{2}w_{-1}^2)- d\xi_t \, w_{-1}.
$$
The corresponding ordinary differential equation for $g_t \equiv
e^{\xi_t w_{-1}}\cdot f_t$ is
\begin{eqnarray*}
 \frac{dg_t(z)}{dt} & = & g_t(z)\frac{1+g_t(z)e^{i\xi_t}}{1-g_t(z)
  e^{i\xi_t}} +2\sum_{m> 0}
\frac{g_t(z)^2e^{i\xi_t}}{e^{2m(p-t)}-g_t(z)e^{i\xi_t}}\\ & &
\hspace{3.17cm}+ 2\sum_{m< 
  0}\frac{g_t(z)e^{2m(p-t)}}{e^{2m(p-t)}-g_t(z)e^{i\xi_t}}. 
\end{eqnarray*}

This time we have, alas, yet no natural group to interpret $\gamma_{f
  _t}$ or $\gamma_{g
  _t}$, because no point is fixed under the evolution.

At time $t=t_c \equiv p$, the evolution terminates because the hull
touches the inner boundary of the annulus, and the distribution of the
hitting point is one of the first problems that comes to mind. This
question is related to the behavior of the solutions of the stochastic
differential equation that describes the motion of points on the inner
boundary $|Z|=e^{-p+t}$ at time $t$. 
Setting $f_t=e^{\zeta_t}$, the stochastic differential equation
eq.(\ref{eqdiff}) becomes
$$d\zeta_t = id\xi_t+dt\left(1-2\sum_{m\geq 0}
  \frac{e^{\zeta_t}}{e^{\zeta_t}-e^{2m(p-t)}}-2 \sum_{m< 0}
  \frac{e^{2m(p-t)}}{e^{\zeta_t}-e^{2m(p-t)}}\right)$$
The motion of
points on the outer and inner boundary are different.  The motion of
points on the outer boundary $|Z|=1$ is obtained by imposing
$\zeta_t=i\theta_t$ with real $\theta_t$. This is consistent with the
stochastic differential equation and leads to
$$d\theta_t = d\xi_t+dt\left(\cot \frac{\theta_t}{2}+2\sum_{m>
    0}\frac{\sin \theta_t}{\cosh 2m(p-t) -\cos \theta_t}\right).$$
The
motions of points on the inner boundary $|Z|=e^{t-p}$ is obtained by
setting $\zeta_t=t-p+i\varphi_t$. This is again consistent with the
stochastic differential equation and leads to
$$d\varphi_t = d\xi_t+2dt\sum_{m\geq 0}\frac{\sin \varphi_t}{\cosh
  (2m+1)(p-t) -\cos \varphi_t}.$$
The $dt$ coefficient in the
equations for $\theta_t$ and $\varphi_t$ becomes singular when
$t\rightarrow t_c=p$.  
These motions are closely related to the elliptic Calogero Hamiltonian, but not
identical because of the explicit time dependence. 

To conclude this discussion of SLE avatars, let us mention that from
the original point of view of Loewner deterministic evolutions in the
annular case, it may be more natural to work with a Loewner
uniformizing map that fixes a point on the outer circle of the
annulus, say $1$, to describe a growing hull starting from the inner
circle. The corresponding vector field is of course again given by
some automorphic form, but slightly different from our $w_{-2}$.
However, in that case there is no automorphism left, so our formalism
based on symmetries gives no clue to define a Brownian source in a
natural way. There is probably a nice SLE in that situation too, so
that a more unifying idea is needed to get a hand on general SLEs.

\section{CFTs of chordal SLEs.}
\label{sec:CFTchorSLE}

We now describe the relation between SLE and conformal field theory
(CFT) which we developed in ref.\cite{BaBe}.
Another approach has been presented in ref.\cite{WernerFrie}.

Recall that $f_t(z)
\equiv g_t(z) - \xi_t$ satisfies the stochastic
differential equation
$$ 
d f_t = \frac{2dt}{f_t}-d\xi_t.
$$
According to section (\ref{sec:chordal}), to $f_t$
we can associate $\gamma_{f_t}\in N_-$, with $N_-$ the group of germs
of holomorphic functions at $\infty$ of the form $z+\sum_{m\leq -1}
f_mz^{m+1}$. By It\^o's formula, it satisfies:
\begin{eqnarray}
\gamma_{f_t}^{-1} \cdot d \gamma_{f_t}=dt(\frac{2}{z}\partial_z
+\frac{\kappa}{2}\partial_z^2)-d\xi_t\partial_z.
\label{itoN-}
\end{eqnarray} 

The operators $\ell_n=-z^{n+1}\partial_z$ are represented in conformal
field theories by operators $L_n$ which satisfy the Virasoro algebra
$\mathfrak{vir}$, see eq.(\ref{viral}).
We need to recall a few basic facts concerning the Virasoro algebra
and its highest weight representations.  These representations possess
a highest weight vector $\ket{h}$ are such that $L_n\ket{h}=0$ for
$n>0$ and $L_0\ket{h}=h\ket{h}$. The parameter $h$ is called
the conformal dimension of
the representation. Define
$$h_{r;s}={[(r\kappa-4s)^2-(\kappa-4)^2]}/{16\kappa}$$ for
$$c=1-6{(\kappa-4)^2}/{4\kappa}.$$
Values of $h_{r;s}$, $r,s=1,2, \cdots$ label degenerate highest
weight  representations  of the Virasoro algebra, with a singular
vector at level $rs$. 
Many interesting conformal weights that appear in CFTs of SLEs are
related to simple integral or half integral values of $r$ and $s$.

The representations of $\mathfrak{vir}$ are not automatically
representations of $N_-$, one of the reasons being that the Lie
algebra of $N_-$ contains infinite linear combinations of the $\ell_n$'s.
However, as we shall explain in section \ref{sec:conftrans}, highest weight
representations of $\mathfrak{vir}$ can be extended in such a way that
$N_-$ get embedded in a appropriate completion
$\overline{\mathcal{U}(\mathfrak{n}_{-})}$ of the enveloping algebra
of some sub-algebra $\mathfrak{n}_{-}$ of $\mathfrak{vir}$.  This
will allows us to associate to any $\gamma_f\in N_-$ an operator $G_f$
acting on appropriate representations of $\mathfrak{vir}$ and
satisfying $G_{g \circ f}=G_f G_g$.  Implementing this construction for
$f_t$ yields random operators $G_t\equiv G_{f_t}\in
\overline{\mathcal{U}(\mathfrak{n}_{-})}$ which satisfy the
stochastic It\^o equation \cite{BaBe}:
\begin{eqnarray}
 G_{t}^{-1} d G_{t}=dt(-2L_{-2}+\frac{\kappa}{2}
L_{-1}^2)+d\xi_t L_{-1}.
\label{labelle}
\end{eqnarray}  
Compare with eq.(\ref{itoN-}).
This may be viewed as defining a Markov process in 
$\overline{\mathcal{U}(\mathfrak{n}_{-})}$.

Since $G_{t}$ turns out to be the operator intertwining the
conformal field theories in $\mathbb{H}$ and in the random domain
$\mathbb{H}_t$, the basic observation which allows us to couple CFTs
to SLEs is the following \cite{BaBe}: 

\vspace{.3cm}
 {\it 
Let $\ket{\omega}\equiv\ket{h_{1;2}}$ 
be the highest weight vector in the irreducible
highest weight representation of $\mathfrak{vir}$ of central charge
$c_{\kappa}=\frac{(6-\kappa)(3\kappa-8)}{2\kappa}$ and conformal
weight $h_{\kappa}\equiv h_{1;2}=\frac{6-\kappa}{2\kappa}$.

Then $G_{t}\ket{\omega}$ is a local martingale.

Assuming appropriate boundedness conditions on $\bra{v}$, the scalar
$\bra{v}G_{t}\ket{\omega}$ is a martingale so that
${\bf E}[\bra{v}G_{t}\ket{\omega}|\{{G_{u}}\}_{u\leq s}]$
is time independent for $t\geq s$ and: } 
\begin{eqnarray}
{\bf E}[\,\bra{v}G_{t}\ket{\omega}\,|\{G_{u}\}_{u\leq s}]
=\bra{v}G_{s}\ket{\omega}
\label{maineq}
\end{eqnarray}

This result is a direct consequence of eq.(\ref{labelle}) and the null
vector relation $(-2L_{-2}+\frac{\kappa}{2}L_{-1}^2)\ket{\omega}=0$
so that $dG_{t}\ket{\omega}=G_{t}L_{-1}\ket{\omega}d\xi_t$.

This result has the following consequences.  Consider CFT correlation
functions in $\monht$. They can be computed by looking at the same
theory in $\monh$ modulo the insertion of an operator representing the
deformation from $\monh$ to $\monht$, see the next section.  This
operator is $G_{t}$. Suppose that the central charge is
$c_{\kappa}$ and the boundary conditions are such that there is a
boundary changing primary operator of weight $h_{1;2}$ inserted at
the tip of $\monK_t$. Then in average the correlation functions of the
conformal field theory in the fluctuating geometry $\monht$ are time
independent and equal to their value at $t=0$.

\begin{figure}[htbp]
  \begin{center}
    \includegraphics[width=0.9\textwidth]{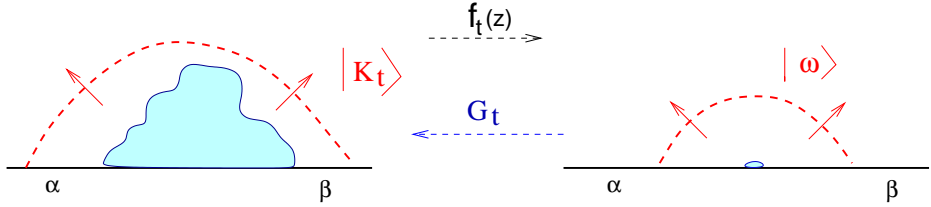}
      \caption{\em A representation of the boundary hull state and of the 
        map intertwining different formulations of the CFT.}
     \label{fig:hilbert}
 \end{center}
\end{figure}

The state $G_{t}\ket{\omega}$ may be interpreted as follows.
Imagine defining the conformal field theory in $\monh_t$ via a radial
quantization, so that the conformal Hilbert spaces are defined over
curves topologically equivalent to half circles around the origin. Then, 
the SLE hulls manifest themselves as disturbances localized around
the origin, and as such they generate states in the conformal Hilbert
spaces. Since $G_{t}$ intertwines the CFT in $\monh$ and in
$\monh_t$, these states are $G_{t}\ket{\omega}$ with
$\ket{\omega}$ keeping track of the boundary conditions.
See Figure (\ref{fig:hilbert}).

\section{CFTs of other SLEs.}

The degenerate representation with highest weight state $\ket{\omega}$
plays a fundamental role in radial, dipolar and annular SLE as well,
and we would like to see this explicitly in this short section. 

In all cases we can choose the geometry in
such a way that the real axis is a boundary globally fixed by the SLE
evolution and that $w_{-1}$ and $w_{-2}$ where meromorphic in a
neighborhood of the origin, with computable small $z$ expansion. The
embedding $\ell_n \rightarrow L_n$ allows to turn $w_{-1}$ and $w_{-2}$
into operators $W_{-1}$ and $W_{-2}$. In all cases, $\ket{\omega}$
turns out to be an eigenvector of the diffusion operator
$\mathcal{A}=-2W_{-2}+\frac{\kappa}{2}W_{-1}^2$. Let us see this explicitly.

\textit{Radial SLE} : We have seen that  $w_{-1}=\ell_{-1}+\ell_1$ and
$w_{-2}=\ell_{-2}+\ell_0$, so that 
$$W_{-1}=L_{-1}+L_1 \qquad W_{-2}=L_{-2}+L_0.$$
A simple rearrangement leads to 
$$-2W_{-2}+\frac{\kappa}{2}W_{-1}^2=\left(-2L_{-2}+\frac{\kappa}{2}L_{-1}^2\right)
  +(\kappa -2)L_0+\kappa L_{-1}L_1+\frac{\kappa}{2}L_{1}^2,$$
Finally
$$\mathcal{A}\ket{\omega}=\frac{(\kappa
  -2)(6-\kappa)}{2\kappa}\ket{\omega}=8h_{0;1/2}\ket{\omega}.$$ 
If $G_t$ represents the action of $\gamma_{f_t}$ in the radial case we
find that
$$M_t \equiv e^{-8h_{0;1/2}t}G_t\ket{\omega}$$
is a local martingale. The prefactor $e^{-8h_{0;1/2}t}$ accounts for
the insertion of a bulk conformal field of scaling dimension $2h_{0;1/2}$
localized at the fixed point, see ref.\cite{BBrad} for further details. 

\textit{Dipolar SLE} : We have seen that  $w_{-1}=\ell_{-1}-\ell_1$ and
$w_{-2}=\ell_{-2}-\ell_0$, so that 
$$W_{-1}=L_{-1}-L_1 \qquad W_{-2}=L_{-2}-L_0.$$
and by the same rearrangement
$$\mathcal{A}\ket{\omega}=-\frac{(\kappa
  -2)(6-\kappa)}{2\kappa}\ket{\omega}=-8h_{0;1/2}\ket{\omega}.$$ 
If $G_t$ represents the action of $\gamma_{f_t}$ in the dipolar case
(we have given two natural candidates) we
find that
$$
M_t \equiv e^{8h_{0;1/2}t}G_t\ket{\omega}$$
is a local martingale.
The prefactor $e^{+8h_{0;1/2}t}$ accounts for the insertion of two
boundary conformal fields, each of dimension $h_{0;1/2}$,
localized at the two fixed points which are the end points of the
forbidden zone.

\textit{Annular SLE} : To see the relation with the Virasoro algebra, we go to a more
convenient geometry.  The change of variable $z=i\frac{1-Z}{1+Z}$ maps
the annulus to the complement of a disk in the upper-half plane
$\monh$. In the new variable one obtains
$w_{-1}=-\frac{1}{2}(1+z^2)\partial_z$, and
$$ w_{-2}^{[p]}=-\frac{1}{4}(1+z^2)\left(\frac{1}{z}+2z\sum_{m> 0}
  \frac{1}{z^2 \cosh^2 mp +\sinh^2 mp}\right)\partial_z.$$ 
The time change to recover the standard normalization is not so
convenient here because of the explicit dependence on the modulus,
hence on time.
Expansion in powers of $z$ reveals that
$w_{-1}=\frac{1}{2}(\ell_{-1}+\ell_1)$ and
$$w_{-2}^{[p]}= \sum_{n \geq -1} a_n \ell_{2n}=\frac{1}{4}\ell_{-2}+
\frac{1}{4}\left(1+2\sum_{m >0} \frac{1}{\sinh^2 mp}\right)\ell_0
+\cdots,$$
with
$$a_n=\frac{(-)^{n}}{2}\sum_{m >0}\frac{\cosh^{2n-2} mp}{\sinh^{2n+2} mp}, \; n
\geq 1.$$
Hence
$$W_{-1}=\frac{1}{2}(L_{-1}+L_1) \qquad
W_{-2}^{[p]}=\frac{1}{4}L_{-2}+ \frac{1}{4}\left(1+2\sum_{m >0}
  \frac{1}{\sinh^2 mp}\right)L_0+\cdots$$
where the missing terms
annihilate $\ket{\omega}$. So :
$$\mathcal{A}^{[p]}\ket{\omega}=\left(2h_{0;1/2}-h_{1;2}\sum_{m >0}
  \frac{1}{\sinh^2 mp}\right)\ket{\omega}.$$
The new feature is that
the eigenvalue is time dependent : it involves the
second Eisenstein series in elliptic function theory.  

For the annular case, we have not constructed explicitly the spaces
$O$ and $N$ so the existence of $\gamma_{f_t}$ remains hypothetical.
 From a physical viewpoint however, the existence of the action of
$G_t$ in conformal field theory makes little doubt, and
$$M_t \equiv \frac{e^{-2h_{0;1/2}t}}{\prod_{m>0}\left(1-e^{2m(t-p)
    }\right)^{2h_{1;2}}}\; G_t\ket{\omega}$$
should be a local martingale. For
large $p$, we retrieve the radial formula. The prefactor is closely
related to the classical partition function, or equivalently to the
Dedekind $\eta$ function. Its modular properties allow to rewrite
$$M_t=e^{-2h_{0;1/2}t} \left(\frac{(p-t)^{\frac{1}{2}}
    e^{-\frac{p-t}{12}+\frac{\pi^2}{12(p-t)}}}
  {\pi^{\frac{1}{2}}\prod_{m>0}\left(1-e^{\frac{2m\pi^2}{t-p}
      }\right)}\right)^{2h_{1;2}}G_t\ket{\omega},$$
which gives the limiting behavior when $t \rightarrow t_c=p$.

\section{Conformal transformations in CFT and applications.}
\label{sec:conftrans}

Recall that we are using boundary CFT on $\monh_t$, the complement
of the SLE hulls, to probe these hulls. We thus need to know how to
implement algebraically conformal maps in CFT.

The basic principles of conformal field theory state that
correlation functions in a domain $\monu$ are
known once they are known in a domain $\monu _0$ and an explicit conformal
map $f$ from $\monu$ to $\monu _0$ preserving boundary conditions is given.
Primary fields have a very simple behavior under conformal
transformations: for a bulk primary field $\Phi$ of weight
$(h,\overline{h})$,
$\Phi(z,\overline{z})dz^hd\overline{z}^{\overline{h}}$ is
invariant, and for a boundary conformal field $\Psi$ of weight
$\delta$, $\Psi(x)|dx|^{\delta}$ is invariant. Their statistical
correlations in $\monu $ and $\monu _0$ are related by
\begin{eqnarray}
 & &\statav{\cdots \Phi(z,\overline{z})\cdots
   \Psi(x)\cdots}_{\monu} 
\nonumber\\
&& \hskip -1.5 truecm
=\statav{\cdots \Phi(f(z),\overline{f(z)})f'(z)^h
  \overline{f'(z)}^{\overline{h}}\cdots
  \Psi(f(x))|f'(x)|^{\delta}\cdots}_{\monu_0}. 
\label{cftconf}
\end{eqnarray}

Infinitesimal deformations of the underlying geometry are implemented
in local field theories by insertions of the stress-tensor.  In
conformal field theories, the stress-tensor is traceless so that it has
only two independent components, one of which, $T(z)$, is holomorphic
(except for possible singularities when its argument approaches the
arguments of other inserted operators).  The field $T(z)$ itself is not a
primary field but a projective connection,
$$\statav{\cdots T(z) \cdots}_\monu= \statav{\cdots
  T(f(z))f'(z)^2+\frac{c}{12}\mathrm{S}f(z)\cdots}_{\monu _0},$$
with $c$ the CFT central charge and
$\mathrm{S}f(z)= \left(\frac{f''(z)}{f'(z)}\right)'-\frac{1}{2}
\left(\frac{f''(z)}{f'(z)}\right)^2$ 
the Schwarzian derivative of $f$ at $z$.

This applies to infinitesimal deformations of the upper half plane.
Consider an infinitesimal hull $\monK_{\epsilon;\mu}$, whose boundary
is the curve $x\to\epsilon\,\pi\mu(x)$, $x$ real and $\epsilon\ll1$,
so that $\monK_{\epsilon;\mu}=\{z=x+iy\in \monh,\ 
0<y<\epsilon\,\pi\mu(x)\}$.  Assume for simplicity that
$\monK_{\epsilon;\mu}$ is bounded away from $0$ and $\infty$.  Let
$\monh_{\epsilon;\mu}\equiv \monh\setminus\monK_{\epsilon;\mu}$.  To
first order in $\epsilon$, the uniformizing map onto $\monh$ is
$$
z+ \epsilon\,\int_{\mathbb{R}}\frac{\mu(y)dy}{z-y} + o(\epsilon).
$$ 
To first order in $\epsilon$ again,
correlation functions in $\monh_{\epsilon; \mu}$
 are related to those in $\monh$ by insertion of $T$:
\begin{eqnarray}
&& \frac{d}{d\epsilon}
\statav{\left(\cdots \Phi(z,\overline{z})\cdots
  \Psi(x)\cdots\right)}_{\monh_{\epsilon; \mu}} \Big\vert_{\epsilon=0^+} 
\nonumber \\
&& \hskip -1.5 truecm
=-\int_{\mathbb{R}} dy \mu(y)\, 
\statav{ T(y)\left(\cdots \Phi(z,\overline{z})\cdots
  \Psi(x)\cdots\right)}_{\monh}
\label{Tdeform}
\end{eqnarray}
With the basic CFT relation \cite{BPZ} between the stress tensor and
the Virasoro generators, $T(z)=\sum_n L_nz^{n-2}$, this indicates that
infinitesimal deformations of the domains are described by insertions
of elements of the Virasoro algebra.

Finite conformal transformations are implemented in conformal field
theories by insertion of operators, representing some appropriate
exponentiation of insertions of the stress tensor. Let
$\mathbb{U}$ be conformally equivalent to the upper half plane $\monh$
and $f$ the corresponding uniformizing map.  Then, following
\cite{BBpart}, the finite conformal deformations that leads from
the conformal field theory on $\mathbb{U}$ to that on $\mathbb{H}$ can
be represented by an operator $G_f$:
\begin{eqnarray}
\statav{\cdots \Phi(z,\overline{z}) \cdots \Psi(x) \cdots }_\mathbb{U}=
 \statav{ G_f ^{-1}\left(\cdots {\Phi}(z,\overline{z})\cdots 
  {\Psi}(x) \cdots \right) G_f}_{\monh}.
\label{utoh}
\end{eqnarray}
This relates correlation
functions in $\mathbb{U}$ to correlation functions in $\mathbb{H}$ where the
field arguments are taken at the same point but conjugated by $G_f$.
Radial quantization is implicitly assumed in eq.(\ref{utoh}).
Compare with eq.(\ref{cftconf}).

The following is a summary, extracting the main steps, of a
construction of $G_f$ described in details in \cite{BBpart}.
To be more precise, we need to distinguish cases depending whether $f$
fixes the origin or the point at infinity.  We also need a few simple
definitions.  We let $\mathfrak{vir}$ be the Virasoro algebra
generated by the $L_n$ and $c$, and $\mathfrak{n}_{-}$ (resp.
$\mathfrak{n}_{+}$) be the nilpotent Lie sub-algebra of
$\mathfrak{vir}$ generated by the $L_n$'s, $n <0$ (resp. $n >0$), and
by $\mathfrak{b}_{-}$ (resp.  $\mathfrak{b}_{+}$) the Borel Lie
sub-algebra of $\mathfrak{vir}$ generated by the $L_n$'s, $n \leq 0$
(resp. $n \geq 0$) and $c$. We denote by
$\overline{\mathcal{U}(\mathfrak{n}_{-})}$ 
(resp. $\overline{\mathcal{U}(\mathfrak{n}_{+})}$) appropriate
completion of the enveloping algebra of $\mathfrak{n}_{-}$
(resp. $\mathfrak{n}_{+}$). We shall only consider highest weight
vector representations of the Virasoro algebra.

\subsection{Finite deformations fixing $0$.}

Let $N_+$ be the space of power series of the form $z+\sum_{m \geq 1}
f_m z^{m+1}$ which have a non vanishing radius of convergence. With
words, $N_+$ is the set of germs of holomorphic functions at the
origin fixing the origin and whose derivative at the origin is $1$. In
applications to the chordal SLE, we shall only need the case when the
coefficients are real. But it is useful to consider the $f_m$'s
as independent commuting indeterminates.

$N_+$ is a group for composition.  Our aim is to construct a group
(anti)-isomorphism from $N_+$ with composition onto a subset
$\mathcal{N}_+ \subset \overline{\mathcal{U}(\mathfrak{n}_{+})}$ with
the associative algebra product.
 
We let $N_+$ act on $O_0$, the space of germs of holomorphic functions
at the origin, by $\gamma_f\cdot F\equiv F \circ f$ for $f \in N_+$
and $F \in O_0$. This representation is faithful and $\gamma_{g\circ
  f}=\gamma_{f} \gamma_{g}$. We need to know how $\gamma_f$ varies
when $f$ varies as $f\to f+\varepsilon v(f)$ for small $\varepsilon$
and an arbitrary vector field $v$.  Taking $g=z+\varepsilon v(z)$ in
the group law leads to $\gamma_{f+\varepsilon v(f)}F=\gamma_{f}\cdot F
+\varepsilon \gamma_{f}\cdot(v \cdot F) +o(\varepsilon)$, where $v
\cdot F(z) \equiv v(z)F'(z)$ is the standard action of vector fields
on functions.  Using Lagrange inversion formula to determine the
vector field $v$ corresponding to the variation of the indeterminate
$f_m$ yields:
$$
\gamma_f^{-1}\,\frac{\partial
  \gamma_f}{\partial f_m}= \sum_{n\geq m} 
\left(\oint_0 \frac{dw}{2i\pi} 
w^{m+1}\frac{f'(w)}{f(w)^{n+2}}\right)\, z^{n+1}\partial_z.
$$ 
This system of first order partial differential equations makes sense
in $\overline{\mathcal{U}(\mathfrak{n}_{+})}$ if we replace
$\ell_n=-z^{n+1}\partial_z$ by $L_n$. So, we define a connection $A_m$ in
$\overline{\mathcal{U}(\mathfrak{n}_{+})}$ by
\begin{eqnarray*}
A_m(f) \equiv  \sum_{n\geq m}
L_n\, \left(\oint_0 \frac{dw}{2i\pi} w^{m+1}\frac{f'(w)}{f(w)^{n+2}}\right)
 \label{eq:zerocurv}
\end{eqnarray*}
which by construction satisfies the zero curvature condition, 
$\frac{\partial A_l}{\partial f_k}-\frac{\partial A_k}{\partial
  f_l}=[A_k,A_l].$

We may thus construct an element $G_f \in \mathcal{N}_+\subset
\overline{\mathcal{U}(\mathfrak{n}_{+})}$ 
for each $f \in N_+$ by solving the system 
\begin{equation}
  \label{eq:pdesys}
\frac{\partial G_f}{\partial f_m}= -G_f A_m(f), \qquad m \geq 1.  
\end{equation}
This system is guaranteed to be compatible, because $N_+$ is convex
and the representation of $N_+$ on $O_0$ is well defined for finite
deformations $f$, faithful and solves the analogous system.  The
existence and uniqueness of $G_f$, with the initial condition $G_{f=z}=1$,
is clear and the group (anti)-homomorphism property,
$G_fG_g=G_{g\circ f}$, is true because it is true infinitesimally and
$N_+$ is convex.  To lowest orders:
$G_f=1-f_1L_1+\frac{f_1^2}{2}(L_1^2+2L_2)-f_2L_2+\cdots.$

The element $G_{f}$, acting on a highest weight representation of the
Virasoro algebra, is the operator which implements the conformal map
$f$ in conformal field theory.  It acts on the stress tensor by
conjugation as:
\begin{equation}
\label{eq:conjt}
G_f^{-1}T(z)G_f= T(f(z))f'(z)^2+\frac{c}{12}Sf(z),
\end{equation} 
a formula which makes sense as long as $z$ is in the disk of
convergence of $f'(z)$ and $Sf(z)$, but which can be extended by
analytic continuation if $f(z)$ allows it.  A similar formula would
hold if we would have considered the action of $G_f$ on local fields. In
particular, by eq.(\ref{eq:conjt}), $G_f$ induces an automorphism of the
Virasoro algebra by $L_m\to L_m(f)\equiv G_f^{-1}L_mG_f$ with
$G_f^{-1}T(z)G_f=\sum_mL_m(f)z^{-m-2}$:
\begin{eqnarray}
  \nonumber
G_f^{-1}L_mG_f = \frac{c}{12}\,\left( 
\oint_0 \frac{dw}{2i\pi} w^{m+1}Sf(w)\right)
+\sum_{n\geq m}L_n\, \left( \oint_0 \frac{dw}{2i\pi} w^{m+1} 
\frac{f'(w)^2}{f(w)^{n+2}}\right).
\end{eqnarray}

Eq.(\ref{eq:pdesys}), which specifies the variations of $G_f$,
can be rewritten in a maybe more familiar way involving the stress
tensor. Namely, if $f$ is changed to $f+\delta f$ with $\delta
f=\varepsilon v(f)$, then:
$$
\delta G_f =- \varepsilon\, G_f\, \oint_0 \frac{dz}{2i\pi}\,T(z)\,v(z).
$$
If $v$ is not just a formal power series at the origin, but a convergent
one in a neighborhood of the origin, we can freely deform contours in
this formula, thus making contact with the infinitesimal deformations
considered in eq.(\ref{Tdeform}).

\subsection{Finite deformations fixing $\infty$.}

All the previous considerations could be extended to the case
in which the holomorphic functions fix $\infty$ instead of
$0$. Let $N_-$ be the space of power series of the form $z+\sum_{m
  \leq -1} f_{m} z^{m+1}$ which have a non vanishing radius of
convergence. We let it act on $O_{\infty}$, the space of germs of
holomorphic functions at infinity, by $\gamma_f\cdot F\equiv F \circ
f$. The adaptation of the previous computations shows that
$\gamma_f^{-1}\frac{\partial \gamma_f}{\partial f_m}= \sum_{n\leq m}
\left(\oint_{\infty} \frac{dw}{2i\pi} w^{m+1}\frac{f'(w)}{f(w)^{n+2}}\right)
z^{n+1}\partial_z.$ We transfer this relation to
$\overline{\mathcal{U}(\mathfrak{n}_{-})}$ to define an (anti)-isomorphism
from $N_-$ to $\mathcal{N}_-
\subset\overline{\mathcal{U}(\mathfrak{n}_{-})}$ mapping $f$ to $G_f$ such
that
$$\frac{\partial
  G_f}{\partial f_m}= -G_f \sum_{n\leq m} L_n\, 
\left( \oint_{\infty} \frac{dw}{2i\pi}
w^{m+1}\frac{f'(w)}{f(w)^{n+2}}\right), \qquad m \leq -1.$$

\subsection{Dilatations and translations.}

We have been dealing with deformations around $0$ and
$\infty$ that did not involve 
dilatation at the fixed point: $f'(0)$ or $f'(\infty)$ was unity.
To gain some flexibility we may also 
authorize dilatations, say at the origin. The operator
associated to a pure dilatation $z\to\lambda z$ is $\lambda^{-L_0}$. 
One can view a general $f$ fixing $0$ as the composition
$f(z)=f'(0)(z+\sum_m f_mz^{m+1})$ of a deformation at $0$ with
derivative $1$ at $0$ followed by a dilatation, so that
$G_f=G_{f/f'(0)}\, f'(0)^{-L_0}.$ 

We may also implement translations. Suppose that
$f(z)=f'(0)(z+\sum_m f_mz^{m+1})$ is a generic invertible germ of
holomorphic function fixing the origin. If $a$ is in
the interior of the disk of convergence of the power series expansion
of $f$ and $f'(a)\neq 0$, we may define a new germ $f_a(z)\equiv
f(a+z)-f(a)$ with the same properties. The operators $G_f$ and
$G_{f_a}$ implementing $f$ and $f_a$ are then related by  
$G_{f_a}=e^{-aL_{-1}}\,G_f\,e^{f(a)L_{-1}}.$

\vskip .3 truecm

\begin{figure}[htbp]
  \begin{center}
    \includegraphics[width=0.7\textwidth]{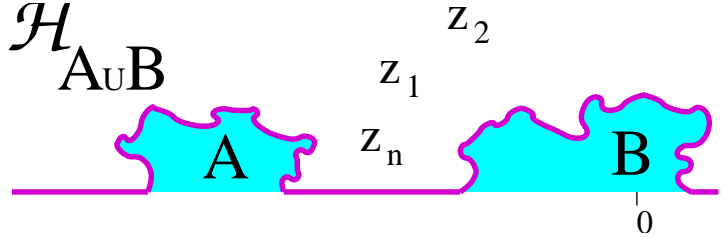}
      \caption{\em A typical two hull geometry.}
      \label{fig:sleblob}
  \end{center}
\end{figure}

\subsection{Deformations around $0$ and $\infty$ 
and `Virasoro Wick theorem'.}

Consider now a domain $\monh_{A\cup B}$ of the type represented on
Figure(\ref{fig:sleblob}) which is the complement of two disjoint
hulls, the first one, say $A$, located around infinity but away from
the origin, and the second one, say B, located around the origin and
away from infinity.  The uniformizing map $f_{A\cup B}$ of
$\monh_{A\cup B}$ onto $\monh$ then does not exist at $0$ or at
$\infty$.

However, in this situation, we may obtain the map $f_{A\cup B}$ by
first removing $B$ by $f_B$, which is regular around $\infty$ and such
that $f_B(z)=z+O(1)$ at infinity, and then $\tild{A} \equiv f_B(A)$ by
$f_{\tild{A}}$ which is regular around $0$ and fixes $0$
($f_{\tild{A}}'(0)\neq 1$ is allowed).  
Of course, the roles of $A$ and $B$ could be interchanged, and we
could first remove $A$ by $f_A$ which is regular around $0$ and fixes
$0$ and then $\tild{B} \equiv f_A(B)$ by $f_{\tild{B}}$ which is
regular around $\infty$ and such that $f_{\tild{B}}(z)=z+O(1)$.

Suppose that $f_A$ and $f_B$ are given. There is some freedom in the
choice of $f_{\tild{A}}$ and $f_{\tild{B}}$~: namely we can replace
$f_{\tild{A}}$ by $h_0 \circ f_{\tild{A}}$ where $h_0\in
PSL_2(\mathbb{R})$ fixing $0$, and $f_{\tild{B}}$ by
$h_{\infty} \circ f_{\tild{B}}$ where $h_{\infty}\in
PSL_2(\mathbb{R})$ such that $h_{\infty}(z)=z+O(1)$ at 
infinity, i.e. a translation. A simple computation shows that
generically there is a unique choice of $f_{\tild{A}}$ and
$f_{\tild{B}}$ such that $f_{\tild{B}}\circ f_A=f_{\tild{A}}\circ
f_B$ and both equal to $f_{A\cup B}$.
This commutative diagram was introduced in \cite{restrictLSW}.

For sufficiently disjoint hulls $A$ and $B$, as in
Figure(\ref{fig:sleblob}), there exists an open set such that for $z$ 
in this set, both products
$G^{-1}_{f_{\tild{A}}}\left(G^{-1}_{f_B}T(z)G_{f_B}\right)G_{f_{\tild{A}}}$
and $G^{-1}_{f_{\tild{B}}}\left(G^{-1}_{f_A}T(z)G_{f_A}\right)G_{f_{\tild{B}}}$
are well defined, given by absolutely convergent series, and are both equal
to $T(f_{A\cup B}(z))f'_{A\cup B}(z)^2+\frac{c}{12}Sf_{A\cup B}(z)$.
As the modes $L_n$ of $~T(z)$ generate all states in a highest weight
representation, the operators $G_{f_B}G_{f_{\tild{A}}}$ and
$G_{f_A}G_{f_{\tild{B}}}$ have to be proportional: they differ at
most by a factor involving the central charge $c$. We write
$G_{f_B}G_{f_{\tild{A}}}=Z(A,B)\; G_{f_A}G_{f_{\tild{B}}},$ or \cite{BBpart}
\begin{equation}
  \label{eq:wik}
  G_{f_A}^{-1}\,G_{f_B}=Z(A,B)\, G_{f_{\tild{B}}}\,G_{f_{\tild{A}}}^{-1}.
\end{equation}

Formula (\ref{eq:wik}) plays for the Virasoro algebra the role that
Wick's theorem plays for collections of harmonic oscillators.
Since $G_{f_A}$ and $G_{f_{\tild{A}}}$ belong to $\mathcal{N}_+$ while 
$G_{f_B}$ and $G_{f_{\tild{B}}}$ to $\mathcal{N}_-$, eq.(\ref{eq:wik})
may also be viewed as defining a product between elements in
$\mathcal{N}_+$ and $\mathcal{N}_-$. Note that
$G_{f_{\tild{B}}}\,G_{f_{\tild{A}}}^{-1}$ is clearly well defined in
highest weight vector representations of $\mathfrak{vir}$.

As implicit in the notation, $Z(A,B)$ depends only on $A$ and $B$: a
simple computation shows that it is invariant if $f_A$ is replaced by
$h_0 \circ f_A$ and $f_B$ by $h_\infty \circ f_B$.  It may be
evaluated as follows. Let $A_s$ and $B_t$ be two families of hulls
that interpolate between the trivial hull and $A$ or $B$ respectively
and $f_{A_{s}}$ and $f_{B_{t}}$ be their uniformizing map. We arrange
that $f_{A_{s}}$ and $f_{B_{t}}$ satisfy the genericity condition, so
that unique $f_{A_{s,t}}$ and $f_{B_{t,s}}$ exist, which satisfy
$f_{B_{t,s}} \circ f_{A_s}=f_{A_{s,t}} \circ f_{B_{t}}$. Define vector
fields by $v_{A_{s}}$ and $v_{B_{t}}$ by $\frac{\partial
  f_{A_{s}}}{\partial s}=v_{A_{s}}(f_{A_{s}})$ and $\frac{\partial
  f_{B_{t}}}{\partial t}=v_{B_{t}}(f_{B_{t}})$. Then \cite{BBpart}:
\begin{eqnarray}
\log Z(A_{\sigma},B_{\tau}) & = & \frac{c}{12}\int_0 ^{\sigma} ds
\oint \frac{dw}{2i\pi} \; v_{A_{s}}(w) Sf_{B_{\tau,s}}(w)\nonumber \\ 
& = & -\frac{c}{12}
\int_0 ^{\tau} dt \oint \frac{dz}{2i\pi} \; v_{B_{t}}(z)
Sf_{A_{\sigma,t}}(z).  
\label{parpart}
\end{eqnarray}
$Z(A,B)$ may physically be interpreted as the interacting part of the
CFT partition function in $\monh \setminus (A\cup B)$.

\section{Deformations and Virasoro representations.}

The above formula may be used to define generalized coherent state
representations of $\mathfrak{vir}$. The key point is to interpret
the `Virasoro Wick theorem', eq.(\ref{eq:wik}), as defining an action
of $\mathfrak{vir}$ on $\mathcal{N}_-$. This is a reformulation of a
construction \`a la Borel-Weil presented in ref.\cite{BBvir}.

\subsection{Representations around infinity.}

Consider first the action of infinitesimal conformal transformations
on $N_-$, i.e. on functions $f(z)=z+\sum_{m \leq -1} f_m z^{m+1}\in
N_-$.  The generator $\ell_{-n}=-z^{1-n}\partial_z$, $n>0$ acts on $N_-$
by $\tild\delta_nf(z)=-z^{1-n}\, f'(z)$. If $n\leq0$, this infinitesimal
variation does not preserve the required behavior of $f$ at infinity
and $z^{1-n}\, f'(z)$ does not belong to $N_-$ for $n\leq 0$.  We may
however restore this behavior by using the commutative diagram
$\tild \phi \circ f = \tild f \circ \phi$ with 
$\phi(z)=z+\varepsilon z^{1-n}$. Then $\tild \phi(z)=z+\varepsilon
\tild y_n(z)$, with $\tild y_n$ polynomial in $z$ of degree $1-n$, and
$\tild f= f + \varepsilon\tild\delta_n f$ belongs to $N_-$ with 
\begin{eqnarray}
\tild\delta_nf(z) = -z^{1-n}\, f'(z) + \tild y_n(f(z))
\label{varsource}
\end{eqnarray}
The polynomial $\tild y_n$ is uniquely fixed by demanding that 
$\tild \delta_nf(z)=o(z)$ at
infinity. Namely, $\widetilde y_n(w)=\sum_ky_kw^{k+1}$ with 
$y_k=\oint_\infty \frac{dz}{2i\pi}z^{1-n}f'(z)^2/f(z)^{k+2}$.

Eqs.(\ref{varsource}) now define an action of the Virasoro algebra,
without central extension, on $N_-$. They have a simple interpretation:
they are infinitesimal conformal transformations in
the source space generated by $\ell_{-n}=-z^{1-n}\partial_z$ preserving the
normalization at infinity.

To get an action of the Virasoro algebra with central extension, 
we  have to slightly generalize this construction using the `Virasoro
Wick theorem'.
Consider a Verma module $V(c,h)$ and take $x\neq 0$ its highest weight
vector. Let $f=z+\sum_{m \leq -1} f_m z^{m+1}\in N_-$
and $G_f$ be the corresponding element in $\mathcal{N}_-$.
The space $\{P_y[f]\equiv \left< y,G_fx \right>, y \in V(c,h)^*\}$, or
$\{Q_y[f]\equiv \langle y,G_f^{-1}x \rangle, y \in V(c,h)^*\}$,
is the space of all polynomials in the independent variables
$f_{-1},f_{-2},\cdots$.  So we have two linear isomorphisms from $V(c,h)^*$
to $\mathbb{C}[f_{-1},f_{-2},\cdots]$ and we can use them
to transport the action of $\mathfrak{vir}$.  We denote by
$\mathcal{R}_n$ and $\mathcal{S}_n$ the differential operators such
that
$$ 
\left<L_ny, G_fx \right>=\mathcal{R}_n \left< y,G_fx \right>
\quad ,\quad 
\langle L_ny,G_f^{-1}x \rangle=
\mathcal{S}_n \langle y,G_f^{-1}x \rangle
$$
for $y \in V^*(c,h)$. By construction the operators $\mathcal{R}_n$ and
$\mathcal{S}_n$ are first order differential operators satisfying the
Virasoro algebra with non vanishing central charge,
$$
[\mathcal{R}_n,\mathcal{R}_m]=(n-m)\mathcal{R}_{n+m}
+\frac{c}{12}n(n^2-1)\delta_{n;-m}, 
$$ 
such that $\mathcal{R}_n\cdot 1=0$ for $n>0$ and 
$\mathcal{R}_0\cdot 1=h$. 
Similar results hold for the $\mathcal{S}_n$'s.
Explicit expressions of these differential operators may be found in
ref.\cite{BBvir, BBpart}.

To be a bit more precise, let us consider $P_y[f]=\left< y,G_fx
\right>$ and $P_{L_ny}[f]=\mathcal{R}_nP_y[f]$.  
We have $\left< L_ny,G_fx\right>=
\left<y,L_{-n}G_fx\right>$. If $n>0$,  $L_{-n}\in \mathfrak{n}_-$,
and, by the group law, the product $L_{-n}G_f$
corresponds to the infinitesimal variation of $\tild\delta_nf(z)$ generated by 
$\ell_{-n}=-z^{1-n}\partial_z$.
If $n\leq0$, $L_{-n}\in \mathfrak{b}_+$ so that we need to re-order the
product $L_{-n}G_f$ in such way that it corresponds to an action
of $\mathfrak{b}_+$ associated to a variation of $f$.
This may be done using the Virasoro Wick theorem,
$G^{-1}_\phi G_f = Z(\phi,f)\, G_{\tild f}G_{\tild \phi}^{-1}$,
eq.(\ref{eq:wik}), which follows from the commutative diagram
$\tild \phi \circ f = \tild f \circ \phi$ with  
$\phi(z)=z+\varepsilon z^{1-n}$ as above.
This diagram  shows that $\phi\in N_+$ acts on $N_-$ by $f\to \tild f$.
For $\phi(z)=z+\varepsilon z^{1-n}$, we have 
$G^{-1}_\phi=1+\varepsilon L_{-n}$ and $G_{\tild
  \phi}^{-1}=1+\varepsilon (G_f^{-1}L_{-n}G_f)_{\mathfrak{b}_+}$
with $(G_f^{-1}L_nG_f)_{\mathfrak{b}_+}=\sum_k y_kL_k$.
The partition function is $Z(\phi,f)=1+\varepsilon c\zeta$ with
$\zeta=\frac{1}{12}\oint_\infty \frac{dz}{2i\pi} z^{1-n} Sf(z)$.
As a consequence, 
$$\left< L_ny,G_fx\right>=(c\zeta+h y_0)\left<y,G_fx\right> 
+ \frac{d}{d\varepsilon}\langle y,G_{\tild f}x\rangle_{\varepsilon=0}.$$
This completely specifies the differential operators $\mathcal{R}_n$.

A similar construction may be used to deal with 
$Q_y[f]=\langle y, G_f^{-1}x \rangle$ 
giving formul\ae\ for $Q_{L_ny}[f]$ as a first order differential
operator $\mathcal{S}_n$ acting on $Q_y[f]$.
Once again the key point is that eq.(\ref{eq:wik}) allows to induce an 
action of $\mathfrak{vir}$ on $\mathcal{N}_-$. The operators
$\mathcal{S}_n$ and $\mathcal{R}_n$ are of course related as one goes
from ones to the others by changing $f$ into its inverse.
As a consequence the variation $f\to \widehat
f=f+\varepsilon \widehat \delta_n f$ induced by $L_n$ is now:
\begin{eqnarray}
\widehat \delta_n f(z)= f(z)^{1-n} - \widehat y_n(z)f'(z)
\label{vartarget}
\end{eqnarray}
where $\widehat y_n(z)$ is uniquely fixed by demanding that $\widehat
\delta_n f(z)=o(z)$, ie. $\widehat y_n(z)= (f(z)^{1-n}/f'(z))_+$ is
the polynomial part of $(f(z)^{1-n}/f'(z))$.  Again,
eqs.(\ref{vartarget}) have a simple interpretation: they are
infinitesimal conformal transformations in the target space generated
by $\widehat \ell_{-n}=f^{1-n}\partial_f$ preserving the normalization at
infinity.

The construction of the differential operators $\mathcal{S}_n$ then
goes as that of the $\mathcal{R}_n$.
In particular, $\mathcal{S}_1$ corresponds to the variation
$\widehat \delta_1f=1$ and $\mathcal{S}_2$ to $\widehat \delta_2f=1/f$:
\begin{eqnarray*}
\mathcal{S}_1 & = & \frac{\partial}{\partial f_{-1}} \\
\mathcal{S}_{2} & = & \sum_{m\leq -2} 
\left(\oint_{\infty} dz\frac{1}{f(z)z^{m+2}}\right)
\frac{\partial }{\partial f_m}.  
\end{eqnarray*}
The other operators $\mathcal{S}_n$ may easily be found,
and are explicitly given in ref.\cite{BBvir, BBpart}.

\subsection{Representations around the origin.}

The presentation parallels quite closely the case of deformations
around $\infty$ so we shall not give all the details. 
Let $f=z+\sum_{m \geq 1} f_m z^{m+1}$ be an element of $N_+$.
Consider a Verma module $V(c,h)$ and take $x$ its highest
weight vector. The space $\{\left< G_fy,x \right>, y \in
V(c,h)^*\}$, or $\{\langle G_f^{-1}y,x \rangle, y \in V(c,h)^*\}$,
 is the space of all polynomials in the independent variables
$f_1,f_2,\cdots$.  So we again have two linear
isomorphisms from $V(c,h)^*$ to $\mathbb{C}[f_1,f_2,\cdots]$, 
and we can use them to transport the
action of $\mathfrak{vir}$. This yields differential operators 
$\mathcal{P}_n$ and $\mathcal{Q}_n$ in the indeterminates $f_m$ such that:
$$
\left< G_fL_ny,x\right>=\mathcal{P}_n \left< G_fy,x \right>
\quad,\quad
\langle G_f^{-1}L_ny,x \rangle
=\mathcal{Q}_n \langle G_f^{-1}y,x \rangle
$$
for $y \in V^*(c,h)$. By construction the operators $\mathcal{P}_n$,
and $\mathcal{Q}_n$, satisfy the Virasoro algebra with central charge
$c$. Their expressions are given in \cite{BBpart}. It is
interesting to notice the operators $\mathcal{Q}_n$, $n\geq 0$, coincide
with those introduced in matrix models. However, the above construction
provides a representation of the complete Virasoro algebra, with
central charge, and not only of one of its Borel sub-algebras.

\section{Applications to SLE.}

The first application is a justification of eq.(\ref{labelle})
which is essential in establishing contact between SLE and CFT.

The SLE maps $g_t$ and $f_t=g_t-\xi _t$ that uniformize the
SLE hull $\monK _t$ fix the point at infinity, so that there are
well defined elements $ G_{g_t}, G_{f_t} \in
\mathcal{N}_-\subset \overline{\mathcal{U} (\mathfrak{n}_{-})}$
implementing them in CFT. The maps are related by a change of the
constant coefficient in the expansion around $\infty$, and the operators
are related by $G_{f _t}=
G_{g_t}e^{\xi _t L_{-1}}$. The map $g_t$ satisfies the
ordinary differential equation $\partial_t
g_t(z)=\frac{2}{g_t(z)-\xi_t}$ and the corresponding vector field
is $v(g)=\frac{2}{g-\xi_t}$, so that
$$G_{g_t}^{-1}dG_{g_t}
=-2e^{\xi _t L_{-1}}L_{-2}e^{-\xi _t L_{-1}}dt.$$
To get $G_{f_t}^{-1}dG_{f_t}$ it remains only
to compute the It\^o derivative of $e^{\xi _t L_{-1}}$ which reads
$e^{-\xi _t L_{-1}}de^{\xi _t L_{-1}}=L_{-1}d\xi
_t+\frac{\kappa}{2}L_{-1}^2dt$.
Finally, 
$$G_{f_t}^{-1}dG_{f_t}=(-2L_{-2}+\frac{\kappa}{2}
L_{-1}^2)dt+ L_{-1}d\xi_t$$
as announced in eq.(\ref{labelle}) with $G_t\equiv G_{f_t}$.

\subsection{Crossing probabilities.}
Crossing probabilities are probabilities associated to some stopping
time events. They have been initially computed by Lawler, Schramm and
Werner using probabilistic arguments \cite{LSW}.
The approach we have been developing \cite{BaBe} relates
them to CFT correlations. It consists in projecting, in an appropriate
way depending on the problem, the martingale equation,
eq.(\ref{maineq}), which, as is well known, may be extended to
stopping times.  Given an event $\mathcal{E}$ associated to a stopping
time $\tau$, we shall identify a vector $\bra{v_\mathcal{E}}$ such
that
$$\bra{v_\mathcal{E}}\, G_{\tau}\ket{\omega}= {\bf
  1}_\mathcal{E}.
$$
The martingale property of $G_{t}\ket{\omega}$ then implies a
simple formula for the probabilities:
$$
{\bf P}[\mathcal{E}]\equiv {\bf E}[\, {\bf 1}_\mathcal{E}\,] = 
\langle{v_\mathcal{E}}\ket{\omega}.
$$

For most of the considered events $\mathcal{E}$, the vectors
$\bra{v_\mathcal{E}}$ are constructed using conformal fields.  The
fact that these vectors  satisfy the appropriate
requirements, $\bra{v_\mathcal{E}}\, G_{\tau}\ket{\omega}= {\bf
  1}_\mathcal{E}$, is then linked to operator product expansion
properties \cite{BPZ} of conformal fields.  This leads to express the
crossing probabilities in terms of correlation functions of 
conformal field theories defined over the upper half plane.
 
Consider for instance Cardy's crossing probabilities \cite{cardy}.
The problem may be formulated as follows.  Let $a$ and $b$ be two
points at finite distance on the real axis with
$a<0<b$ and define stopping times $\tau_a$ and
$\tau_b$ as the first times at which the SLE trace
$\gamma_{[0,t]}$ touches the intervals 
$(-\infty,a]$ and $[b,+\infty)$ respectively.
The generalized Cardy's probability is the probability that
the SLE trace hits first the interval $(-\infty,a]$, that is
${\bf P}[\,\tau_a<\tau_b\,]$. For this event, the vector
$\bra{v_\mathcal{E}}$ is constructed using the product of two
boundary conformal fields $\Psi_0(a)$ and $\Psi_0(b)$ each of
conformal weight $0$. This leads to the formula for $4<\kappa<8$: 
$$
{\bf P}[\,\tau_a<\tau_b\,]=
\frac{C_0(a/b)-C_0(\infty)}{C_0(0)-C_0(\infty)}
$$
where $C_0$ is the CFT correlation function, 
which only depends on $a/b$:
$$ C_0(a/b)=\bra{\omega}\Psi_0(a)\Psi_0(b)\ket{\omega}.$$
More detailed examples have been described in \cite{BaBe}.
Zig-zag probabilities to be discussed below give another illustration
of the method.

Our approach and that of Lawler, Schramm and Werner,
refs.\cite{LSW,schramm0}, are linked but they 
are in a way reversed one from the other.  Indeed, the latter evaluate
the crossing probabilities using the differential equations they
satisfy -- because they are associated to martingales,-- while we
compute them by identifying them with CFT correlation functions --
because they are associated to martingales -- and as such they satisfy
the differential equations.

\subsection{The restriction martingale.}

Let us now go to another application to SLEs by giving a CFT
interpretation of the restriction martingale first computed in
ref.\cite{restrictLSW}.  
As already mentioned the basic point is eq.(\ref{maineq}) 
which says that $G_{t}\ket{\omega}$ is a local martingale.

We apply the results of the previous two hull construction in the
case when $B$ is the growing SLE hull $\monK_t$ and $A$ is another
disjoint hull away from $\monK_t$ and the infinity. Let $f_A$ be the
uniformizing map of $\monh\setminus A$ onto $\monh$ fixing the origin.
Since $G_{t}\ket{\omega}$ is a local martingale, so is
$M_A(t)\equiv\aver{G_{f_A}^{-1}G_{t}}$.

To compute it, we start from $f_A$ and $f_t$ 
to build a commutative diagram as in previous section,
with maps denoted by $f_{\tild{A}_t}$ and $\tild{f}_t$
uniformizing respectively $f_t(A)$ and $f_A(\monK_t)$ and satisfying
$\tild{f}_t \circ f_A=f_{\tild{A}_t} \circ f_t$. Then, from
eq.(\ref{eq:wik}) with $G_t\equiv  G_{f_t}$, we have:
$$
\aver{G_{f_A}^{-1}G_{t}} =
  Z(A,\monK_t)\,\aver{G_{\tild{f}_t}\,G_{f_{\tild{A}_t}}^{-1}}
$$
$Z(A,\monK_t)$ may be computed using eq.(\ref{parpart}):
$\log Z(A,\monK_t)={-\frac{c}{6}\int_0^{t} d\tau Sf_{A_{\tau}}(0)}.$
We have
$\aver{G_{\tild{f}_t}\,G_{f_{\tild{A}_t}}^{-1}}=
f_{\tild{A}_t}'(0)^{h_{\kappa}}.$
Thus the partition function martingale $M_A(t)$ reads:
$$M_A(t)=f_{\tild{A}_t}'(0)^{h_{\kappa}}\ \exp {-\frac{c_\kappa}{6}\int_0
  ^{t} d\tau\,  Sf_{A_{\tau}}(0)}.$$ 

This local martingale was discovered without any recourse to
representation theory in \cite{restrictLSW} but
it is clearly deeply rooted in CFT. From it, one
may deduce \cite{restrictLSW} the probability that, for $\kappa=8/3$,
the SLE trace $\gamma_{[0,\infty]}$ does not touch $A$:
$${\bf P}[\gamma_{[0,\infty]}\cap A=\emptyset]= f_{A}'(0)^{5/8}$$
where $f_A$ has been further normalized by $f_A(0)=0$ and
$f_A(z)=z+O(1)$ at infinity.
Recall that for $\kappa=8/3\leq 4$, the SLE hull $\monK_t$ coincides
with the SLE trace $\gamma_{[0,t]}$ and that it almost surely avoids
the real axis at any finite time.

\subsection{Generating martingale algebra.}
We may also  rephrase the main result, eq.(\ref{maineq}),
in more algebraic language, see ref.\cite{BBvir}. 
Let $\mathcal{R}_n$ and $\mathcal{S}_n$ be the
differential operators define above and consider $f_t$ the SLE
map. Its coefficients $f_{-1},f_{-2},\cdots$ are random (for instance
$f_{-1}$ is simply a Brownian motion of covariance $\kappa$).  Because
$\mathcal{S}_n$, $n> 0$, are the differential operators implementing
the variation $\widehat \delta_n f=f^{1-n}$, 
the stochastic Loewner evolution
(\ref{loew}) may be written in terms of the Virasoro generators
$\mathcal{S}_n$ acting on functions $\mathcal{F}[f]$ of the $f_m$:
$$
d\mathcal{F}[f]
=dt\, (2\mathcal{S}_{2}+\frac{\kappa}{2}\mathcal{S}_{1}^2)\mathcal{F}[f]
- d\xi_t\,\mathcal{S}_{1}\mathcal{F}[f]
$$

Consider now the Verma module $V(c_\kappa,h_\kappa)$, with
$c_\kappa=\frac{(6-\kappa)(3\kappa-8)}{2\kappa}$ and
$h_\kappa=\frac{6-\kappa}{2\kappa}$. It is not irreducible, since
$(-2L_{-2}+\frac{\kappa}{2}L_{-1}^2)x$ is a singular vector in
$V(c_\kappa,h_\kappa)$, annihilated by the $L_n$'s, $n \geq 1$, so
that it does not couple to any descendant of $x^*$, the dual of $x$.
The descendants of $x^*$ in $V^*(c_\kappa,h_\kappa)$ generate the
irreducible highest weight representation of weight
$(c_\kappa,h_\kappa)$.  If $y$ is a descendant of $x^*$,
$\left<y,G_f(-2L_{-2}+\frac{\kappa}{2}L_{-1}^2)x \right>=0$, 
or equivalently,
$$ (2\mathcal{S}_{2}+\frac{\kappa}{2}\mathcal{S}_{1}^2)\left<y, G_fx
\right>=0$$
since, as function of the $f_m$, $\left<y, G_fL_{-n}x\right>
=-\mathcal{S}_n\left<y, G_fx \right>$ for $n \geq 1$.

As a consequence, all the
polynomials in $f_{-1},f_{-2},\cdots$ obtained by acting repeatedly on
the polynomial $1$ with the $\mathcal{R}_m$'s (they build the
irreducible representation with highest weight $(c_\kappa,h_\kappa)$) are
annihilated by $2\mathcal{S}_{2}+\frac{\kappa}{2}\mathcal{S}_{1}^2$.
For generic $\kappa$ there is no other singular vector in
$V(c_\kappa,h_\kappa)$, and this leads to a satisfactory description
of the irreducible representation of highest weight
$(c_\kappa,h_\kappa)$: the representation space is given by the
kernel of an explicit differential operator acting on
$\mathbb{C}[f_{-1},f_{-2},\cdots]$, and the states are build by repeated
action of the differential operators $\mathcal{R}_m$ on
the highest weight state $1$.

The above computation can be interpreted as follows:
the space of polynomials of the coefficients of the expansion
of $f_t$ at $\infty$ for SLE$_\kappa$ can be endowed with a
Virasoro module structure isomorphic to $V^*(c_{\kappa},h_{\kappa})$.
Within that space, the subspace of (polynomial) martingales is a submodule
isomorphic to the irreducible highest weight representation of weight
$(c_{\kappa},h_{\kappa})$. 

\section{Boundary zig-zag probabilities.}

What we call boundary zig-zag probabilities are the probabilities that
the SLE curve visits a set of intervals on the real axis in a given
order.  We are going to show on a few examples how these probabilities
are related to particular CFT correlation functions of boundary
primary fields. These relations reveal connections between topological
properties of SLE paths and fusion algebras and operator production
expansions in conformal field theory.

In this section we assume $4<\kappa<8$ and 
we shall use freely known results from conformal field theory.

We shall use the following notation.
For $I_p=[x_p,X_p]$ disjoint intervals on the real axis ($|x_p|<|X_p|$,
$x_pX_p>0$) let  
$$
P_n(I_1;\cdots;I_n)\equiv {\bf P}[\tau_{x_1}<\tau_{X_1},\cdots,
\tau_{x_n}<\tau_{X_n}\ {\rm and}\
\tau_{x_1}<\tau_{x_2}<\cdots <\tau_{x_n}]
$$
be the probabilities that SLE curves touch these intervals in the
order used to index them, i.e. $I_1,\ I_2,\cdots, I_n$.

\subsection{0ne interval probabilities.}

Consider the probability $P_1([x,X])$ that the SLE curve touches an
interval $[x,X]$ on the positive real axis. In terms of swallowing
times this is the probability ${\bf P}[\tau_x<\tau_X]$.  It is
directly related to the distribution of the position $a_0$ of the
first point on the real axis bigger than $x$ touched by the SLE path.
Indeed the probability that $a_0\geq X$ is simply the probability that
$x$ and $X$ are swallowed at the same instant, ie. the probability
that $\tau_x=\tau_X$. Hence,
$$ P_1([x,X])\equiv {\bf P}[\tau_x<\tau_X] = 1-  {\bf P}[\tau_x=\tau_X]$$
We shall compute ${\bf P}[\tau_x=\tau_X]$, which
was already determined in ref.\cite{RhodeSchramm}, 
by relating it to a CFT correlation function with dimension zero
boundary primary fields inserted at end points of the interval
$[x,X]$. 

To prepare for the computation, we study the correlation function
$$\bra{\omega} \Psi_{0}(X)\Psi_{0}(x)\ket{\omega}.$$

If $x$ comes close to $0$, we can expand this function by computing
the operator product expansion of $\Psi_{0}(x)\ket{\omega}$.  This is
constrained by the fusion rules which arise from the null vector
$(L_{-2}-\frac{\kappa}{4}L_{-1}^2)\ket{\omega}=0$.  It can involve at
most two conformal families of dimension $h_{1;2}$ or $h_{1;0}$.  We
demand that only the conformal family of dimension $h_{1;0}$ appears
in the operator product expansion. This fixes the boundary conditions
we impose on the correlation functions.  In other words, we choose the
primary field $\Psi_0(x)$ to intertwines Virasoro modules of
dimensions $h_{1;2}$ and $h_{1;0}$.  Then, $\Psi_{0}(x)\ket{\omega}
\sim x ^{\frac{\kappa -4}{\kappa}} \ket{h_{1;0}}$, and this goes to
$0$ for $\kappa > 4$.

If the points $x$ and $X$ come close together, the operator product expansion 
$\Psi_{0}(X)\Psi_{0}(x)$ is more involved. General rules of
conformal field theory ensure that the identity operator contributes,
but apart from that, there is no a priori restrictions on the
conformal families $\Psi_{\delta}$ that may appear. 
However, only
those for which $\bra{\omega} \Psi_{\delta}\ket{\omega}\neq 0$
remain, and this restricts to two conformal families, the identity and
$\Psi_{1;3}$. Namely, when $x$ and $X$ come close together,
$$
\bra{\omega} \Psi_{0}(X)\Psi_{0}(x)\ket{\omega}\simeq 1+ \cdots +
(X-x)^{h_{1;3}} C_1\, \bra{\omega}\Psi_{1;3}(x)\ket{\omega} +\cdots
$$
with $h_{1;3}=\frac{8-\kappa}{\kappa}$ and $C_1$ some fusion
coefficient.  The dominant contribution to $\bra{\omega}
\Psi_{0}(X)\Psi_{0}(x)\ket{\omega}$ is either $1$ or
$(X-x)^{h_{1;3}}$, depending on whether $\kappa<8$ or $\kappa>8$.

Hence, if $4<\kappa <8$, the correlation function 
$$
\bra{\omega} \Psi_{0}(X){}_{[1;0]}\Psi_{0}(x)\ket{\omega},
$$
vanishes as $x\to 0$ and takes value $1$ at $X\to x$.  
Here, we have specified with the index ${}_{[1;0]}$ that the
intermediate conformal family has dimension $h_{1;0}$.

Now, for nonzero $t$, we write  
\begin{eqnarray*}
\bra{\omega}\Psi_{0}(X){}_{[1;0]}\Psi_{0}(x)G_t\ket{\omega}
& = & \bra{\omega} \Psi_{0}(f_t(X)) {}_{[1;0]}
\Psi_{0}(f_t(x)) \ket{\omega} \\
& = & \bra{\omega} \Psi_{0}(1) {}_{[1;0]}
\Psi_{0}(f_t(x)/f_t(X)) \ket{\omega}.
\end{eqnarray*}
The last equality follows by dimensional analysis.  If $a_0$, the
position of the SLE trace at $t=\tau_x$, satisfies $x< a_0 < X$,
$f_{\tau_x}(X)$ remains away from the origin but $f_{\tau_x}(x)\to 0$ 
and the correlation function vanishes. On the other hand, if $X\leq a_0$,
it is a general property of hulls that $\lim _{t \nearrow
\tau_x}f_t(x)/f_t(X)=1$ and the correlation function is unity.
Thus 
$$\bra{\omega} \Psi_{0}(X){}_{[1;0]}
\Psi_{0}(x) G_{\tau _x}\ket{\omega}=\mathbf{1}_{a_0 \geq X}
=\mathbf{1}_{\tau_x=\tau_X}.$$
 From the martingale property (\ref{maineq}), we infer that
$$
{\bf E}[ \bra{\omega} \Psi_{0}(X){}_{[1;0]}
\Psi_{0}(x) G_{\tau _x}\ket{\omega}]=
\bra{\omega} \Psi_{0}(X){}_{[1;0]}\Psi_{0}(x)\ket{\omega}
$$
and thus
\debut
{\bf P}[\tau_x=\tau_X]=
\bra{\omega}\Psi_{0}(X){}_{[1;0]}\Psi_{0}(x)\ket{\omega}.
\label{excur0}
\fin
This example is instructive, because it shows in a fairly simple case
that the thresholds $\kappa=4,8$ for topological properties of SLE
appear in the CFT framework as thresholds at which divergences emerge
in operator product expansions. It also clearly shows how the intermediate 
conformal families appearing in the CFT correlation functions are
selected according to the topological behavior specified by the
probabilities one computes.

Furthermore, the fact that 
$\bra{\omega}\Psi_{0}(X)\Psi_{0}(x)(-2L_{-2}+
\frac{\kappa}{2}L_{-1}^2)\ket{\omega}=0$ 
translates into a differential equation for
the correlations functions.
Since this correlation function depends only on $s=x/X$,  
we derive that
$$\left(\frac{d^2}{ds^2}+\left(\frac{4}{\kappa s} +
\frac{2(4-\kappa)}{\kappa(1-s)}\right)\frac{d}{ds}
\right)\bra{\omega} \Psi_{0}(1)\Psi_{0}(s)
\ket{\omega}=0.$$
The differential operator annihilates the
constants, a remnant of the fact that the identity operator has weight
0. With the chosen normalization for $\Psi_{0}(x)$, the
relevant solution vanishes at the origin. The integration is
straightforward. Finally, with $s=x/X$,
$${\bf P}[\tau_x=\tau_X]=
\frac{s^{\frac{\kappa -4}{\kappa}}\Gamma\left(\frac{4}{\kappa}
\right)}{\Gamma\left(\frac{\kappa-4}{\kappa}\right)\Gamma
\left(\frac{8-\kappa} {\kappa}\right)}\int_0^1 d\sigma \sigma
^{-\frac{4}{\kappa}} (1-s\sigma)^{2\frac{4-\kappa}{\kappa}}.$$

Recall that the probability that the SLE path touches the interval
$[x,X]$ is simply $P_1([x,X])= 1 - {\bf P}[\tau_x=\tau_X]$.
The limit of an infinitesimal interval may be taken by fusing
$\Psi_0(X)$ and $\Psi_0(x)$ together. We then get 
$(X-x)^{h_{1;3}}C_1\,\bra{\omega}\Psi_{1;3}(x)\ket{\omega}$,
or alternatively,
\begin{eqnarray}
P_1([x,x+dx]) = C_1\ (dx/|x|)^{h_{1;3}} \quad ,\quad  
h_{1;3}=\frac{8-\kappa}{\kappa}
\label{1dense}
\end{eqnarray}
with $C_1$ the fusion constant.
This formula indicates that the operator coding for two SLE paths
emerging from the real axis is the boundary operator $\Psi_{1;3}$.

\subsection{A recursion formula.}

The Markov property of SLE implies a simple recursion formula
for the probabilities $P_n(I_1;\cdots;I_n)$:
\begin{eqnarray}
&& P_{n}(I_1;\cdots;I_{n}) \nonumber\\
&& \hskip -1.5 truecm
={\bf E}[ {\bf 1}_{ \tau_{x_1}<\tau_{X_1} }\
{\bf 1}_{ \tau_{x_1}<\min(\tau_{x_2},\cdots ,\tau_{x_n}) }\
P_{n-1}(f_{\tau_{x_1}}(I_2);\cdots;f_{\tau_{x_1}}(I_{n}) ) ]
\label{zigrecur}
\end{eqnarray}

The Markov property of SLE means that, for $t>s$,
$f_t\circ f_s^{-1}$ and $f_{t-s}$ 
are identically distributed 
and that $f_t\circ f_s^{-1}$ is independent of $f_r$ for $0\leq r <s<t$.

Eq.(\ref{zigrecur}) may be understood 
by considering the event $\mathcal{E}_1$ for
which the SLE curve first touches the interval $I_1$, i.e.
$\mathcal{E}_1=\{\tau_{x_1}<\tau_{X_1},\
\tau_{x_1}<\min(\tau_{x_2},\cdots,\tau_{x_n})\}$.  
We have the standard relation
${\bf E}[{\bf 1}_{\mathcal{E}_1}{\bf 1}_{\mathcal{E}_2}]
={\bf E}[{\bf 1}_{\mathcal{E}_1} {\bf P}[{\mathcal{E}_2}|\mathcal{E}_1]]$
for conditional probabilities.
We look at the probability ${P}_{n-1}[I_2;\cdots;I_{n}|\mathcal{E}_1]$ 
that the SLE curve touches the intervals $I_2;\cdots;I_{n}$ knowing 
that it has first touched the interval $I_1$. 
At the instant $\tau_{x_1}$ at which the SLE curve touches $I_1$, 
the other intervals are mapped into
$f_{\tau_{x_1}}(I_2);\cdots;f_{\tau_{x_1}}(I_{n})$. 
By the Markov property we may restart the SLE process at time $\tau_{x_1}$,
so that the last conditional probability is simply
$P_{n-1}(f_{\tau_{x_1}}(I_2);\cdots;f_{\tau_{x_1}}(I_{n}))$.
Eq.(\ref{zigrecur}) then follows.

The Markov property has another consequence on zig-zag probabilities.
Looking at the evolution of the SLE trace during a short period of
time tells us that the process $t\to P_n(f_t(I_1);\cdots;f_t(I_n))$,
defined as long as none of the interval has been swallowed, is a local
martingale.  By It\^o calculus, this implies differential equations
for $P_n(I_1;\cdots;I_n)$. Supplemented with appropriate boundary
conditions, they may allow to determine $P_n$.  This is however not
the route we would like to follow as it does not reveal the CFT
interpretation of the zig-zag probabilities.

\subsection{An example of zig-zag probabilities.}

\begin{figure}[htbp]
  \begin{center}
    \includegraphics[width=0.6\textwidth]{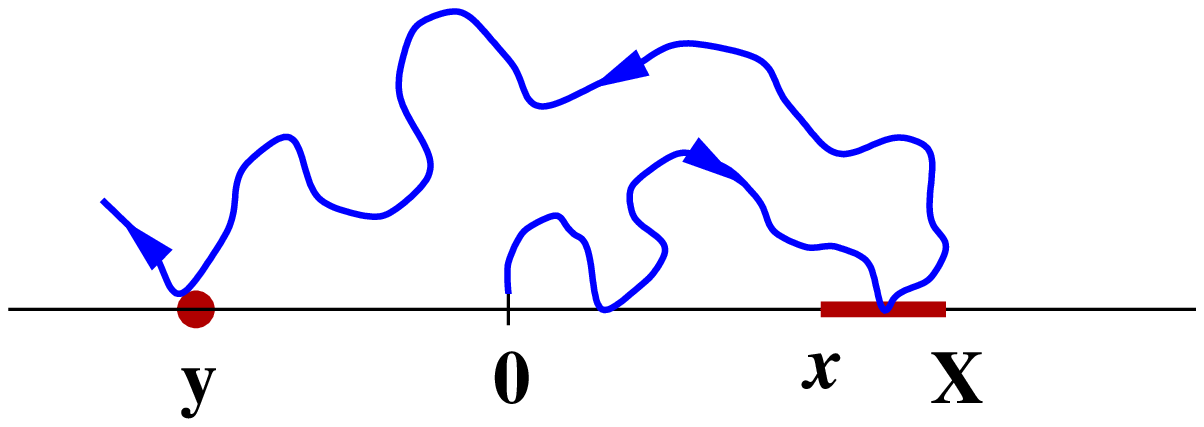}
      \caption{\em Two interval zig-zag probability.}
      \label{fig:zigzag}
  \end{center}
\end{figure}

To generalize further our previous example, let us consider
the zig-zag probability that the SLE traces touch say first an interval
$[x,X]$ on the positive real axis and then an infinitesimal interval
located at $y<0$.  See Figure(\ref{fig:zigzag}).
 From the result of the previous section, we know
that this probability scales as $(dy)^{h_{1;3}}$.  We denote it by
$$P_2([x,X],[y,y+dy])\equiv Q_2(x,X;y)(dy)^{h_{1;3}}.$$
By the recursion formula (\ref{zigrecur}) and eq.(\ref{1dense}), we
have: 
\begin{eqnarray}
Q_2(x,X;y)= C_1\ 
{\bf E}[ {\bf 1}_{ \tau_{x}<\tau_{X} }\
{\bf 1}_{ \tau_{x}<\tau_{y} }\
\Big|{\frac{f'_{\tau_x}(y)}{f_{\tau_x}(y)}}\Big|^{h_{1;3}} ]
\label{1petit1gros}
\end{eqnarray}
By definition, it satisfies the boundary conditions:
\begin{eqnarray} 
Q_2(x,X;y)  \cases{ = 0 ,&  for $y\to 0$;\cr
  \simeq  C_1\, |y|^{-h_{1;3}}+\cdots,& for $x\to 0$, $\forall X$;\cr
 =0 ,&  for $X\to x$;\cr}
\label{condlimit}
\end{eqnarray}
The first one follows from the fact that as $y\to 0$ the SLE trace
swallows the point $y$ before $x$ with probability one. The second one 
is a consequence of the fact that as $x\to 0$ the SLE trace hits first
the interval $[x,X]$ with probability one so that 
$P_2([x,X],[y,y+dy])\simeq P_1([y,y+dy])$ as $x\to 0$. 
The third one is obvious.

\vskip .3 truecm

We shall identify $Q_2(x,X;y)$ with a particular CFT correlation
function which involves a primary field $\Psi_{1;3}$ of weight
$h_{1;3}$ localized on the infinitesimal interval $[y, y +dy]$ and two
primary fields $\Psi_0$ of weight zero localized at the end points of
the macroscopic interval $[x,X]$.

So let us consider the correlation function
\begin{eqnarray}
F(X,x;y)\equiv \bra{\omega} \Psi_0(X) {}_{[1;2]}\, \Psi_0(x) {}_{[1;4]}\,
    \Psi_{1;3}(y) \ket{\omega} 
\label{2intercor}
\end{eqnarray}
The intermediate indices ${}_{[1;2]}$ and ${}_{[1;4]}$ refer to the
conformal weights of the Virasoro representations which propagate in
the intermediate channels, eg. $\Psi_{1;3}(y)$ in the above formula is
the primary field of dimension $h_{1;3}$ intertwining Virasoro modules
of weights $h_{1;2}$ and $h_{1;4}$. By the CFT fusion rules, which are
consequences of the null vector relation
$(L_{-2}-\frac{\kappa}{4}L_{-1}^2)\ket{\omega}=0$, 
in the correlation function (\ref{2intercor}), the intermediate representation
between the dimension zero operators $\Psi_0(X)$ and
$\Psi_0(x)$ could only have weight $h_{1;0}$ or $h_{1;2}$ -- and we
choose $h_{1;2}$ -- whereas the intermediate representation between
$\Psi_0(x)$ and $\Psi_{1;3}(y)$ could only have weight $h_{1;2}$ or
$h_{1;4}$ -- and we choose $h_{1;4}$ there.  Choosing these
representations amounts to select a particular conformal block. 

We are going to argue (i) that, up to a proportionality coefficient,
$F(X,x;y)$ satisfies the same boundary
conditions (\ref{condlimit}) as $Q_2(x,X;y)$ -- and this is how 
we have chosen the intermediate channels, 
and (ii) that, up to the same proportionality coefficient,
it is such that
\begin{eqnarray}
F\Big(f_\tau(X),f_\tau(x);f_\tau(y)\Big)\, |f'_\tau(y)|^{h_{1;3}}
={\rm const.}\, {\bf 1}_{ \tau_{x}<\tau_{X} }\
{\bf 1}_{ \tau_{x}<\tau_{y} }\
\Big|{\frac{f'_{\tau_x}(y)}{f_{\tau_x}(y)} }\Big|^{h_{1;3}}
\label{2interevent}
\end{eqnarray}
for $\tau=\min(\tau_x,\tau_y)$.
This will follow from the fact that $F(x,X;y)$ satisfies
the boundary conditions (\ref{condlimit}). 
Since by the martingale property, we know that
$$ {\bf E}[\, F\Big(f_\tau(X),f_\tau(x);f_\tau(y)\Big)\, 
|f'_\tau(y)|^{h_{1;3}}\, ] = F(X,x;y),
$$
this will imply that  
$F(X,x;y)={\rm const.}\, Q_2(y;x,X)$,
where ${\rm const.}$ is the above mentioned proportionality
coefficient.

Let us first look at the boundary conditions.
Consider the limit $y\to 0$.  By CFT fusion rules, the operator
$\Psi_{1;3}(y)$ maps $\ket{\omega}$ in a Virasoro module of
weight $h_{1;2}$ or $h_{1;4}$. Demanding that
$\Psi_{1;3}(y)\ket{\omega}$ vanishes as $y\to 0$ selects the image
space to be of Virasoro weight $h_{1;4}$, because then
$\Psi_{1;3}(y)\ket{\omega}\simeq y^{4/\kappa} \ket{h_{1;4}}+\cdots$,
without contribution from the $h_{1;2}$ module which would be divergent.
This fixes the intermediate channel between $\Psi_0(x)$ and
$\Psi_{1;3}(y)$ to be ${}_{[1;4]}$.

To fix the other intermediate channel we look at the limit $x\to 0$.
In this limit, $x$ is approaching the localization point of
$\ket{\omega}$, so we have to exchange the order by which we act with 
$\Psi_0(x)$ and $\Psi_{1;3}(y)$ using the CFT braiding relations:
$$
\Psi_0(x) {}_{[1;4]}\,\Psi_{1;3}(y) \ket{\omega} 
= B_{[1;4]}^{[1;2]}\ \Psi_{1;3}(y) {}_{[1;2]}\,\Psi_{0}(x)\ket{\omega} 
+ B_{[1;4]}^{[1;0]\ }\Psi_{1;3}(y) {}_{[1;0]}\,\Psi_{0}(x)\ket{\omega} 
$$
There are only two contributions to this braiding relation because of
the fusion rules implied by the level two null vector 
$(L_{-2}-\frac{\kappa}{4}L_{-1}^2)\ket{\omega}=0$.
In the limit $x\to 0$, the first term dominates and
$\Psi_0(x) {}_{[1;4]}\,\Psi_{1;3}(y) \ket{\omega} 
\simeq B_{[1;4]}^{[1;2]}\ \Psi_{1;3}(y)\ket{\omega} + \cdots$. 
This gives, as $x\to 0$,
$$
\bra{\omega} \Psi_0(X) \Psi_0(x) {}_{[1;4]}\, \Psi_{1;3}(y) \ket{\omega} 
\simeq B_{[1;4]}^{[1;2]}\  \bra{\omega} \Psi_0(X)
\Psi_{1;3}(y)\ket{\omega} + \cdots
$$
The boundary condition (\ref{condlimit}) demands the above r.h.s. be
proportional to $|y|^{-h_{1;3}}$ for all $X$. This fixes the intermediate
representation between $\Psi_0(X)$ and $\Psi_{1;3}(y)$ to be of weight
$h_{1;2}$ and not $h_{1;0}$, because $ \bra{\omega} \Psi_0(X) {}_{[1;2]}
\Psi_{1;3}(y)\ket{\omega}=|y|^{-h_{1;3}}$ for all $X$.

The last boundary condition for $X\to x$ is again a consequence of
the fusion relation of the dimension zero operators: 
$$
\bra{\omega}\Psi_0(X) {}_{[1;2]} \Psi_0(x)  {}_{[1;4]}
\simeq (X-x)^{h_{1;3}}\, \bra{\omega}\Psi_{1;3}(x){}_{[1;4]} +
\cdots
$$
Only the operators $\Psi_{1;3}$ and $\Psi_{1;5}$ may appear in this
fusion, and the $\Psi_{1;3}$ contribution is dominant as $X\to x$.

Let us now show that these boundary conditions imply that $F(X,x;y)$ 
satisfies eq.(\ref{2interevent}). We have to distinguish the three
topologically distinct configurations: (i) $\tau_y<\tau_x\leq\tau_X$,
(ii) $\tau_x=\tau_X<\tau_y$ and (iii) $\tau_x<\tau_y$ and
$\tau_x<\tau_X$. In case (i), point $y$ is swallowed
before $x$ and $X$, thus $\tau=\tau_y$ and $f_\tau(y)\to 0$ with 
$f_\tau(x)$ and $f_\tau(X)$ finite. The first boundary condition
(\ref{condlimit}) then implies that 
$F(f_\tau(X),f_\tau(x);f_\tau(y))\, |f'_\tau(y)|^{h_{1;3}}$ vanishes.
In case (ii), points $x$ and $X$ are swallowed simultaneously before
$y$, so $\tau=\tau_x$ and $(f_\tau(x)-f_\tau(X))\to 0$ faster that 
$f_\tau(x)$ and $f_\tau(X)$ while $f_\tau(y)$ remains finite.
The third boundary condition (\ref{condlimit}) implies that
$F(f_\tau(X),f_\tau(x);f_\tau(y))\, |f'_\tau(y)|^{h_{1;3}}$ also vanishes
in this case. In case (ii), the SLE trace touches the interval $[x,X]$ 
before swallowing $y$, hence $\tau=\tau_x$ and $f_\tau(x)\to 0$ with
$f_\tau(X)$ and $f_\tau(y)$ finite. The second boundary condition
(\ref{condlimit}) then implies that
$F(f_\tau(X),f_\tau(x);f_\tau(y))\, |f'_\tau(y)|^{h_{1;3}}$ is equal to 
$|f'_{\tau_x}(y)/f_{\tau_x}(y)|^{h_{1;3}}$.
This proves eq.(\ref{2interevent}) and hence
$F(X,x;y)={\rm const.}\, Q_2(y;x,X)$.

We can now go on and look for the density probability of hitting two
infinitesimal intervals, say on both side of the origin, 
touching first $[x,x+dx]$ and then $[y, y+dy]$ with $y<0<x$:
$$ P_2([x,x+dx];[y,y+dy] )\equiv 
Q_{2;2}(x;y)\,(dy)^{h_{1;3}}(dx)^{h_{1;3}}$$
This may be obtained from eq.(\ref{2intercor}) by fusing the
weight zero primary operators as above:
$\bra{\omega}\Psi_0(X) {}_{[1;2]}\Psi_0(x){}_{[1;4]} \simeq C_1 (X-x)^{h_{1;3}}\,
\bra{\omega}\Psi_{1;3}(x){}_{[1;4]}+\cdots$. We get, up to a proportionality
coefficient:
$$
Q_{2;2}(x;y) ={\rm const.}\, \bra{\omega}
\Psi_{1;3}(x){}_{[1;4]}\Psi_{1;3}(y)\ket{\omega} 
$$
where, again, the index ${}_{[1;4]}$ specifies the intermediate
channel. It satisfies a 
hypergeometric differential equation and it may thus be written in
terms of hypergeometric function. The explicit formula, although
easy to obtain, is not very useful at this point.
This correlation function may also be identified, 
up to a proportionality coefficient, with
$$
{\bf E}[ {\bf 1}_{\tau_x<\tau_y} \Big|
\frac{f'_{\tau_x}(x)f'_{\tau_x}(y)}{f_{\tau_x}(x)f_{\tau_x}(y)}
\Big|^{h_{1;3}}].
$$
This shows that each infinitesimal interval counts for the insertion 
of a factor $\big|f'_{\tau}(x)/f_{\tau}(x)\big|^{h_{1;3}}$ in the SLE
expectation or for the insertion of the boundary primary field
$\Psi_{1;3}$ in the CFT correlation functions.


This example illustrate how zig-zag probabilities of SLE paths may be
identified with CFT correlation functions. It shows that fusion
algebras of CFT are at the heart of this identification.

Other examples dealing with more intervals, in various side of the
origin and visited by the SLE path in various order, may be
considered.  However, the identification of the appropriate CFT
correlation functions which will code for these zig-zag probabilities
becomes more and more involved. For instance, it is clear that
probabilities that the SLE path touches infinitesimal intervals are
given by correlation functions of $\Psi_{1;3}$ operators and that
specifying the order in which these intervals are visited selects the
appropriate conformal block. But we didn't find any simple rule
encompassing all cases.

\section{Bulk zig-zag probabilities.}
Bulk zig-zag probabilities are probabilities for the SLE curve to
visit a set of balls on the upper half plane in a given order.
The following  is essentially a reformulation, using CFT approach, of 
Beffara's proof \cite{Beffara} that the fractal dimension of a SLE
curve is $d_\kappa=1+\kappa/8$ for $\kappa\leq 8$ and $d_\kappa=2$ for
$\kappa\geq 8$, a formula which was conjectured by Duplantier \cite{Duplan}.

\subsection{Fractal dimension and one-point function.}
The fractal dimension $d_\kappa$
may be evaluated by using a one point estimate, namely 
the probability that the SLE path approaches
a point $z_0$ in the bulk of the upper half plane at a distance less
that $\varepsilon$:
$${\bf P}[\gamma_{[0,+\infty)}\cap
\mathbb{B}_\varepsilon(z_0)\not=\emptyset] 
\approx \varepsilon^{2-d_\kappa} ,\quad \varepsilon\to 0
$$ 
with $\mathbb{B}_\varepsilon(z_0)$ the ball of radius $\varepsilon$
centered in $z_0$. This yields \cite{Duplan,Beffara}:
$d_\kappa= 2 -2 h_{0;1}= 1 + \kappa/8$ for $\kappa<8$.

The complete determination of $d_\kappa$ requires also establishing a
two point estimate, which is much harder to obtain but which may be
found in the nice reference \cite{Beffara}.

Let $z_0\in \mathbb{H}$, $\Im {\rm m}z_0>0$, be a point in the upper
half plane and $\delta_t(z_0)$ its distance to the SLE curve
$\gamma_{[0,t]}$ stopped at time $t$. We shall evaluate $\delta_t(z_0)$
using the conformal radius of $\gamma_{[0,t]}$ seen from $z_0$.
Let $h_t(z)$, defined by
$$ h_t(z)= \frac{g_t(z)-g_t(z_0)}{g_t(z)-\overline{g_t(z_0)}}$$
 be a uniformizing map of
$\mathbb{H}\setminus\mathbb{K}_t$ onto the unit disk with
$h_t(z_0)=0$, $h_t(\infty)=1$. It maps the tip of the curve on the
unit circle, $U_t\equiv h_t(\gamma(t))\in \mathbb{S}_1$, with 
$U_t=f_t(z_0)/\overline{f_t(z_0)}\equiv e^{i\alpha_t}$.
This defines a process $\alpha_t$ on the unit circle with 
$\alpha_t\to 0$ or $2\pi$ as $t\to \tau_{z_0}$ depending whether
$z_0$ has been swallowed clockwise, or counterclockwise, by the SLE
trace. Actually, up to a random time reparametrization,
$ds=(\frac{2\Im {\rm m}f_t(z_0)}{|f_t(z_0)|^2})^2dt$, 
this process is driven by 
$d\alpha_s= \frac{\kappa-4}{4}\cot (\alpha_s/2) + d\xi_s$.

The conformal radius of $\gamma_{[0,t]}$ viewed from $z_0$ is defined
as $\rho_t(z_0)\equiv |h_t'(z_0)|^{-1}$.  An explicit computation
shows that $\rho_t(z_0)=|2\Im {\rm m} f_t(z_0)/f'_t(z_0)|$.  K\"obe
$1/4$-theorem states that $\delta_t(z_0)$ and $\rho_t(z_0)$ scale the
same way, since $(1/4)\rho_t(z_0)\leq \delta_t(z_0)\leq \rho_t(z_0)$.
One may check that $\rho_t(z_0)$ is always decreasing as time goes by.
So instead of estimating the distance between the SLE path and $z_0$,
we shall estimate its conformal radius, $\rho(z_0,\gamma)=\lim_{t\to
  \tau_{z_0}}\rho_t(z_0)$, and the probability
\begin{eqnarray}
 {\bf P}[\rho(z_0,\gamma)\leq \varepsilon ]
\label{1pointestim}
\end{eqnarray}

Estimating ${\bf P}[\rho(z_0,\gamma)\leq \varepsilon ]$ can be
formulated \cite{Beffara} as a survival probability problem for the
process $\alpha_s$ but, in order to understand its CFT origin, we
shall compute it using our favorite martingale $G_t\ket{\omega}$.
For $\kappa<8$, let us consider the expectation value
$$ 
\hat M_t(z_0)\equiv \bra{\omega}\Phi_{0;1}(z_0,\bar z_0)G_t\ket{\omega}
$$
with $\Phi_{0;1}$ the bulk conformal field of weight
$2h_{0;1}=(8-\kappa)/8$. By construction this is well defined up to
time $t<\tau_{z_0}$.  The correlation function
$\bra{\omega}\Phi_{0;1}(z_0,\bar z_0)\ket{\omega}$ may be computed
exactly using the level two null vector. 
It is equal to $$|2\Im{\rm m}z_0|^{-2h_{0;1}}
(\sin\alpha_0/2)^{\kappa/8-1}$$ with $z_0/\bar z_0= e^{i\alpha_0}$ so
that
$$
\hat M_t(z_0)= \Big\vert \frac{f'_t(z_0)}{2\Im{\rm m}f_t(z_0)}
\Big\vert^{2h_{0;1}}\, (\sin\alpha_t/2)^{\kappa/8-1}
= \rho_t(z_0)^{-2h_{0;1}}\, (\sin\alpha_t/2)^{\kappa/8-1}
$$
Let $\theta_{z_0}^\varepsilon$ be either the time at which the
conformal radius $\rho_t(z_0)$ reaches the value $\varepsilon$,
if $\rho(z_0,\gamma)\leq \varepsilon$, or the
swallowing time $\tau_{z_0}$ if the point $z_0$ is swallowed before 
the conformal radius reaches this value, 
i.e. if $\rho(z_0,\gamma)> \varepsilon$. 
The time $\theta_{z_0}^\varepsilon$ is a stopping time.
Since $f'_t(z_0)$ vanishes faster than $f_t(z_0)$ as $t\to
\tau_{z_0}$, the martingale
$\hat M_t(z_0)$ vanishes as $t\to \tau_{z_0}$ for $\kappa<8$.
Therefore it projects on
$$ 
\hat M_{\theta_{z_0}^\varepsilon}(z_0)= \varepsilon^{-2h_{0;1}}\, 
(\sin\alpha_{\theta_{z_0}^\varepsilon}/2)^{\kappa/8-1}\
{\bf 1}_{\rho(z_0,\gamma)\leq \varepsilon}
$$
By construction $\hat M_t(z_0)$ is a martingale so that 
${\bf E}[\hat M_{\theta_{z_0}^\varepsilon}(z_0)]=\hat M_{t=0}(z_0)$, 
or, using the basic definition of conditional probability,
\begin{eqnarray*}
&&{\bf P}[\rho(z_0,\gamma)\leq \varepsilon]\ \cdot \
{\bf E}[(\sin\alpha_{\theta_{z_0}^\varepsilon}/2)^{\kappa/8-1}
\, \vert\, \rho(z_0,\gamma)\leq \varepsilon] \\
&& \hskip 1 truecm
= \Big(\frac{\varepsilon}{2\Im{\rm m}z_0}\Big)^{2h_{0;1}}\
(\sin\alpha_0/2)^{\kappa/8-1}
\end{eqnarray*}
The conditional probability 
${\bf E}[(\sin\alpha_{\theta_{z_0}^\varepsilon}/2)^{\kappa/8-1}
\, \vert\, \rho(z_0,\gamma)\leq \varepsilon]$ is bounded so
that
${\bf P}[\rho(z_0,\gamma)\leq \varepsilon]\approx
\varepsilon^{2h_{0;1}}$ as $\varepsilon\to 0$.
This gives the one point estimate (\ref{1pointestim}) 
and the fractal dimension $d_\kappa= 2 - 2h_{0;1}$.

\subsection{Bulk zig-zags and $N$-point functions.}

The previous computation indicates that the operator testing for the
emergence of two SLE paths from a tiny ball in $\mathbb{H}$ is the
bulk primary field $\Phi_{0;1}$.
\vskip .3 truecm

\begin{figure}[htbp]
  \begin{center}
    \includegraphics[width=0.6\textwidth]{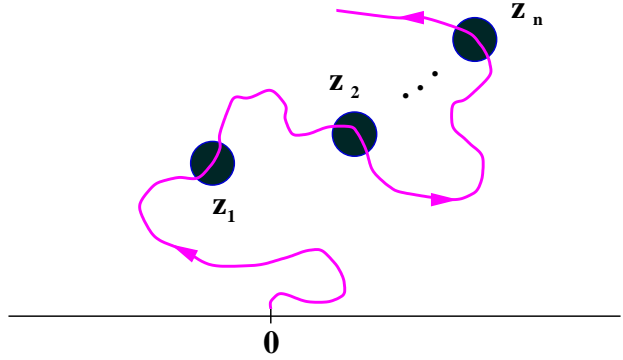}
      \caption{\em Bulk zig-zag probabilities.}
      \label{fig:zbulk}
  \end{center}
\end{figure}

As in Figure(\ref{fig:zbulk}),
one may look for the zig-zag density probability that the SLE path
visit balls $\mathbb{B}_\varepsilon(z_p)$, $\varepsilon \to 0$, 
centered in points $z_p$,
$$
\lim_{\varepsilon\to 0} \varepsilon^{-2nh_{0;1}}\,
{\bf P}[\gamma_{[0,\infty)}\ {\rm visits}\
B_\varepsilon(z_1)\cup\cdots\cup B_\varepsilon(z_n)]
$$
in the order $z_1,\cdots,z_n$. 
This is clearly proportional to a CFT
correlation functions of field of dimension $h_{0;1}$:
$$
\bra{\omega} \Phi_{0;1}(z_1,\bar z_1)\cdots 
\Phi_{0;1}(z_n,\bar z_n) \ket{\omega}
$$
As in the boundary case, to define these correlation functions 
one has to specify the Virasoro representations
propagating in the intermediate channels.

If no order among the visited balls is specified, these correlation
functions have no monodromy and they thus correspond to the complete
$\Phi_{0;1}$ CFT correlation functions.  Zig-zag probabilities
with specified ordering in the visits would probably be exchanged as
one moves the points $z_p$ around. In other words, there is probably 
a quite direct relation between CFT monodromies, alias quantum groups, 
and braiding properties of samples of SLE traces.

\vskip 1.5 truecm

{\bf Acknowledgments:} 

Work supported in part by EC contract number
HPRN-CT-2002-00325 of the EUCLID research training network.

\vskip 1.5 truecm


\end{document}